# Scientific elite revisited:
# Patterns of productivity, collaboration, authorship and impact

Jichao Li[1,2,3,4], Yian Yin[2,3,5], Santo Fortunato[6,7] and Dashun Wang[2,3,4,5*]

[1] *College of Systems Engineering, National University of Defense Technology, Changsha, China*

[2] *Center for Science of Science and Innovation, Northwestern University, Evanston, IL, USA.*

[3] *Northwestern Institute on Complex Systems, Northwestern University, Evanston, IL, USA.*

[4] *Kellogg School of Management, Northwestern University, Evanston, IL, USA*

[5] *McCormick School of Engineering, Northwestern University, Evanston, IL, USA*

[6] *School of Informatics, Computing, and Engineering, Indiana University, Bloomington, IN, USA*

[7] *Indiana University Network Science Institute (IUNI), Indiana University, Bloomington, IN, USA*

\* e-mail: dashun.wang@northwestern.edu

# ABSTRACT

Throughout history, a relatively small number of individuals have made a profound and lasting impact on science and society. Despite long-standing, multi-disciplinary interests in understanding careers of elite scientists, there have been limited attempts for a quantitative, career-level analysis. Here, we leverage a comprehensive dataset we assembled, allowing us to trace the entire career histories of nearly all Nobel laureates in physics, chemistry, and physiology or medicine over the past century. We find that, although Nobel laureates were energetic producers from the outset, producing works that garner unusually high impact, their careers before winning the prize follow relatively similar patterns as ordinary scientists, being characterized by hot streaks and increasing reliance on collaborations. We also uncovered notable variations along their careers, often associated with the Nobel prize, including shifting coauthorship structure in the prize-winning work, and a significant but temporary dip in the impact of work they produce after winning the Nobel. Together, these results document quantitative patterns governing the careers of scientific elites, offering an empirical basis for a deeper understanding of the hallmarks of exceptional careers in science.

# Introduction

According to Harriet Zuckerman[1], scientific elites "are worthy of our attention not merely because they have prestige and influence in science, but because their collective contributions have made a difference in the advance of scientific knowledge". Indeed, across the broad spectrum of sciences, scientific elites are often pathbreakers and pacesetters in the science of their time[2-7]. Understanding patterns governing the careers of scientific elites helps us uncover insightful markers for exceptional scientific careers, useful for scientists and decision makers who hope to identify and develop individual careers and institutions[8].

The Nobel Prize, widely regarded as the most prestigious award in science, offers a unique opportunity to systematically identify and trace many of the world's greatest scientists[1,3,8-15]. These scientific elites have attracted interest from a wide range of disciplines[1,3,8,11,12,15-27], spanning sociology, economics, psychology, and physics. On the one hand, quantitative studies analyzing publication and citation records have mainly focused on the prize-winning work alone, helping uncover a set of highly reproducible patterns ranging from understanding the link between age and creativity[3,16,17,28-30], to allocating credits and recognition[4,15,19,21]. On the other hand, Zuckerman's canonical work[1] probes into the *entire* career histories of Nobel laureates through qualitative methods[13,14,16,31-35]. The rich patterns articulated by Zuckerman vividly highlight the need to go beyond their prize-winning works, and put them in the context of the entire careers of laureates. Together, the two strands of research call for a quantitative, career-level analysis relying on large-scale datasets to study patterns of productivity, collaboration, authorship, and impact governing the careers of scientific elites.

Despite recent surge of interest in the science of science[3,19,28,29,36-43] and efforts in constructing large-scale datasets of scholarly activities[3,44-46], large-scale studies of the career histories of Nobel laureates remained limited, largely owing to the difficulty in collecting systematic data for their scientific contributions. Here, by combining information collected from the Nobel Prize official websites, laureates' university websites, Wikipedia entries, publication and citation records from the Microsoft Academic Graph (MAG)[47], and extensive manual curations, we constructed a unique dataset capturing career histories of nearly all Nobel laureates in physics, chemistry, and physiology or medicine from 1900 to 2016 (545 out of 590, 92.4%)[48]. We cross-validated this dataset with four different approaches to ensure the reliability of our results. We deposited the derived dataset in a public data repository[49], and describe our data collection and validation procedures in a data descriptor with great detail[48].

We further constructed a comparison dataset of scientific careers using data from the Web of Science (WOS) and Google Scholar (GS)[46], representing the kinds of "ordinary" careers that tend to be studied in the science of science literature[29,50]. For each laureate who published the first paper after 1960, we randomly selected 20 scientists in the same discipline who started their careers in the same year (Supplementary Information S1). Note that the goal here is not to create a matching sample of Nobel-caliber scientists, but a comparison group consisting of scientists who are more similar to typical scientists in the field. One advantage of this comparison approach is that, by selecting individuals with long careers and well-maintained GS profiles, it covers scientists with relatively higher visibility and impact than typical scientists, indicating that our comparisons offer a conservative estimate of the difference between Nobel laureates and their contemporary peers.

## Results

**Early Performance.** Widely held is the belief that the great minds do their critical work early in their careers[3,16,17], prompting us to ask if there is any early signal that distinguishes Nobel laureates. Here we focus on the first five years since their first publication and measure their productivity and impact at this early stage of their careers. Consistent with Zuckerman's observation[1], we find that Nobel laureates were energetic producers from the outset, publishing almost twice as many papers as scientists in our comparison group (Fig. 1a). Yet, compared with this productivity difference, more impressive is the gap in impact. Indeed, the future laureates had a more than six-fold increase over the comparison group in terms of the rate of publishing hit papers, defined as the papers with top 1% of rescaled 10-year citations (Eq. (1)) in the same year and field (Supplementary Information S3.1) (Fig. 1b). This difference is not simply driven by the early onset of prize-winning works. Indeed, we repeated our measurements by omitting the careers of laureates who published their prize-winning work in this period, finding that a substantial gap remained (Fig. S1).

To conceptualize the observed difference in productivity and impact, we separated team- and solo-authored papers, finding that both types of work boost early performance, but they do so in different ways: Most of the difference in early productivity is accounted for by team-authored papers, as solo-authored papers show meager productivity difference between the laureates and their comparison group (Fig. 1c), documenting a greater propensity toward collaborations for scientific elites in their early careers[1]. The only exception is physics laureates who published slightly more solo-authored papers than their comparison group (1.73 vs 1.07, Student's t-test, p-value=0.07). Yet interestingly, solo-authored papers in early careers turned out to disproportionally more likely to be prize-winning papers than team-authored ones. Indeed, comparing the fractions of prize-winning papers within solo- and team-authored papers, we find that the former is about twice as high as the latter on average (Chi-squared test, p-value<$10^{-11}$, Fig. 1d).

**Career before the prize.** Figure 1 documents outstanding early performance of future laureates. This is consistent with the innovation literature, which shows that the most important works tend to occur early in the lifecycle[3,16,51], speaking to the idea that great, young minds disproportionally break through. Yet on the other hand, growing evidence shows that ordinary scientific careers are governed by the random impact rule[28], predicting that the highest impact work occurs randomly within the sequence of works. To reconcile these two schools of thought, we focus on the career of laureates before they were awarded the Nobel and measure the positions of the prize-winning work and highest impact work within the sequence of works one produced. Here the paper impact is measured by rescaled 10-year citation (Methods). We find both types of works tend to occur early within the sequence of papers (Fig. 2a), a result that contradicts the random impact rule governing typical scientific careers[28,46]. Yet, our earlier analysis suggests that a selection effect may offer a potential explanation for this observation[52]—since the Nobel Prize in science has never been awarded posthumously, those who produced ground-breaking works early were more likely to wait long enough to be recognized[20,22]. Indeed, we removed prize-winning papers and calculate among the remaining ones the position of the highest impact papers. We find that the timing of each of the three remaining highest impact works for Nobel laureates all follow clearly uniform patterns[52] (Fig. 2b). This means, apart from the prize-winning work, all other important works in Nobel careers closely follow the random impact rule: They could be, with equal likelihood, the very first work, the last, or any one in between. This observation is in line with the recent discovery of hot streaks that occur at random within individual careers[46], and therefore raises an important next question: Are these high-impact works clustered together in time?

To answer this question, we quantify the relative timing between the two most cited papers ($N^*$ and $N^{**}$) within each career by calculating the joint probability $P(N^*, N^{**})$ with a null model in which the two papers each follow their independent temporal patterns. We uncovered clear diagonal patterns across all three domains (Figs. 2c-e), showing that high-impact papers are more likely to cluster together than expected by chance. The diagonal pattern disappeared when we shuffle the order of the works, while preserving the random impact rule (Figs. 2f-h). We also measured the distribution of the longest streak within a career $L$, finding that $P(L)$ follows a broader distribution compared with that in shuffled careers across all three disciplines (Figs. 2i-k) (Supplementary Information S4.3-4.5). We further find that their hot streaks occur randomly within the sequence of works (Fig. 2l), and are not associated with any detectable change in the overall productivity (Fig. 2m, Kolmogorov-Smirnov test, $p$-value=0.18). Together, these results demonstrate a remarkable resemblance between the career histories of Nobel laureates and ordinary scientists[46].

What seems to distinguish the Nobel laureates from ordinary scientists, however, is that they are disproportionately more likely to have more than one hot streak. Indeed, while hot streak is usually unique for typical scientists[46], Nobel laureates are characterized by 1.93 hot streaks on average (Fig. 2n). Furthermore, their hot streaks also tend to sustain for longer. We measured the duration distribution of hot streaks for Nobel laureates, finding that it peaks around 5.2 years (Fig. 2o), compared with 3.7 years for typical scientists[46]. The longer duration of laureates' hot streaks is also captured by its proportion over career length (Fig. 2p). We also find that prize-winning works are disproportionately more likely to be produced during hot streaks (Fig. 2q). Overall the vast majority of all Nobel winning works (88%) occurred within hot streaks.

**Collaboration patterns.** One of the most fundamental shifts in science over the past century is the flourishing of large teams across all areas of science[29,39,53,54]. Compared with the overall rate of this shift, Nobel laureates' papers are produced by an even higher proportion of large teams (Fig. 3a). One possible factor that may explain this team-size difference is impact, as larger teams tend to produce papers with higher impacts[37]. To control for this factor, we created a matching sample for each paper published by the laureates by selecting 20 papers from the same field and year but with the most similar number of citations. We find that, after controlling for impact, the Nobel laureates' papers are still more likely to be produced by larger teams in all times across the last century (Fig. 3b).

Figure 3ab thus underscore another similarity between Nobel and ordinary careers, highlighting the increasing reliance of team work across all types of scientific careers. Yet the ubiquitous increase in team size can be in tension with the fact that the Nobel Prize can only be awarded to at most three recipients for each subject every year[1], prompting us to compare the team size of all prize-winning papers with those published immediately before and after them by the same laureates[52] (Supplementary Information S5.1). We find a greater propensity for the prize-winning papers to be written by less than three authors[52] (61.43% vs 53.28%, Chi-squared test, p-value<$10^{-4}$, Fig. 3c). We further examine the authorship structure of the prize-winning papers, finding that they are substantially more likely to have the laureates as the first author than other joint papers published by them (45.04% vs 30.64%, Chi-squared test, p-value<$10^{-7}$) (Fig. 3d). We also calculated the probability of being the last author, finding no statistical difference (Chi-squared test, p-value=0.41).

To test if these phenomena are unique to the prize-winning work, we removed the prize-winning papers and repeated the same analysis for the most cited paper among the remaining papers. We find that there is no statistical difference in their likelihood of being written by small teams[52] (60.18% vs 56.17%, Chi-squared test, p-value=0.1193, Fig. 3e). While the difference in the likelihood of being the first author still exists for chemistry laureates, there is no statistical difference for laureates in physics or medicine (Fig. 3f). Together, these results show that prize-winning papers are more likely to be authored by fewer than three authors, with an intriguing tendency for laureates to claim the first authorship in the prize-winning works. While these observations are consistent with the finding that works produced by small teams tend to disrupt science and technology[37], they are also consistent with Zuckerman's argument that "the future laureates were especially concerned to have the record clear for their most significant work, and particularly in their prize-winning research papers"[1].

**After the prize.** How does winning the Nobel impact one's subsequent career? The Matthew effect[4,55] tells us that winning begets more winnings. Hence one may expect that works produced after the Nobel garner more impact than those produced before, given their substantially elevated reputation and visibility[15]. Here we find that, to the contrary, when comparing the average impact of papers (defined in Eq. (4)) published by the laureates in each of the four years before and after winning the Nobel, the average impact per paper shows a significant *drop* in the two years following the Nobel. The effect is most significant in the year immediately after, where impact dropped by 11.1% on average compared with the year before. Furthermore, the effect is not permanent, with impact quickly bouncing back by year four to a similar level as the year of the Nobel (Fig. 4a). The "Nobel dip" is most pronounced for physics laureates, as the impacts of their papers were reduced by 18.1%, compared with 4.8% for chemistry and 13.4% for medicine (Supplementary Information S6.2, Fig. S17). Interestingly, in contrast to the common perception of decreased productivity following the Nobel[1,21], possibly due to "the disruptive consequences of abrupt upward social mobility"[1], we find that the average number of papers by the laureates shows no significant change (Fig. 4b), indicating that the uncovered Nobel dip mainly pertains to impact rather than productivity. Note that winning the Nobel may introduce citation boosts to prior papers by the laureate[15,26]. To understand if the observed dip in impact may be explained by this factor, we alter the observation window to exclude post-Prize citations to pre-Prize works, finding that the "dip and bounce back" pattern remains robust (Fig. 4c, Supplementary Information S6.4, Fig. S20). We also find that the number of solo-authored papers decreased precipitately after the Nobel (Student's t-test, *p*-value=0.004, Fig. 4d), whereas the fraction of team-authored papers increased (Student's t-test, *p*-value=0.008, Fig. 4e), suggesting that collaboration and teamwork carry an increasing importance for the laureates after winning the Nobel.

The Nobel dip signals that scientific community's attention is not driven by status but the quality of work. To unearth potential mechanisms underlying the "dip and bounce back" dynamics, we trace topic changes before and after Nobel as reflected in their publications. We use an established method[43] that detects research topics based on communities in the co-citing network of papers published by a scientist, offering a discipline-independent method to identify and trace research topics across a career (Supplementary Information S6.5). As an illustrative example, Fig. 4f shows the constructed co-citing network and topic communities for the career of Jean-Marie Lehn, who was awarded the 1987 chemistry Nobel together with Donald Cram and Charles Pedersen for the synthesis of cryptands. In his remarkable career, Lehn published more than 700 papers. Figure 4f visualizes his publication history by topic, showing that his research agenda was almost exclusively focused on cryptands related research, until he was awarded the Nobel in 1987. Yet, just as this line of research was officially recognized, we observed a clear shift in topic right after winning the Nobel (Fig. 4g). In the next 10 years, his research was primary focused on self-assembly and self-organization. Most interestingly, this is a topic that he had never published on before winning the Nobel.

The intriguing example of Lehn's career prompts us to ask if laureates disproportionately shift research topic after winning the Nobel. We randomly selected two papers, within 4 years before and after the Nobel, respectively, and measured the probability of two papers belonging to the same topic, finding only 36.8% of the two papers cover the same topic before and after winning the Prize. We then build a null model by randomly choosing a year as the pretended prize-winning year for comparison, finding that the probability is significant higher (45.2% vs 36.8%, $p$-value=0.004, Fig. 4h), which suggests the laureates have a higher likelihood of shifting research topics after winning the Nobel. We further measured the likelihood of laureates studying a new topic after winning Prize, and compare it with a null model where we shuffled the topic of the works, finding that the laureates are much more likely to study a new topic after winning Prize than expected (14.2% vs 1.8%, $p$-value<$10^{-14}$, Fig. 4i). To ensure that these results are not affected by specific community detection methods used to detect topics, we repeated our analyses with another well-known algorithm (Infomap[56]), obtaining the same conclusions (Supplementary Information S6.6).

To understand potential forces behind the uncovered change in research agenda following the Nobel, we examined several different factors, including the popularity of research topics before and after the Prize (Supplementary Information S6.7), changes in collaborators (Supplementary Information S6.8), and funding opportunities (Supplementary Information S6.9). We find that the topic studied after the Nobel tends to be less popular at the time. The number of new collaborators does not increase after the Nobel, but these collaborators tend to more established in terms of productivity and impact. And somewhat surprisingly, the overall funding to each laureate remains mostly constant around the time of the award. Although none of these factors can directly explain the observed topic change and the associated citation dip (Figs. 4j-m, Supplementary Information S6.7-S6.9, Figs. S24-S26), they appear consistent with an endogenous shift in the laureate's interest to explore new directions. Note that, although the uncovered dip-bounce-back dynamics and topic shifting behavior both occur around the same time (when awarded the Nobel), it does not imply that the two are causally related. On the other hand, while one may be better at anticipating which work will be recognized by the Nobel eventually[57], it remains difficult to precisely predict the year of winning, indicating that the award year can be viewed as a largely exogenous variation in a career[58], which then coincides with topic shifting behavior that is largely endogenous to the individual. Regardless, these results highlight the unwavering scientific efforts by the laureates, actively pursuing new lines of enquiry while undeterred by the extra burdens imposed by growing duties and responsibilities[1].

**Discussion**

In summary, building on Zuckerman's canonical work on scientific elites[1], here we present a systematic empirical investigation of the careers of Nobel laureates by studying patterns of productivity, collaboration, authorship, and impact. This analysis is now possible thanks to a novel dataset we curated—both algorithmically and manually—which links several disparate biographical and bibliographical data sources, offering a unique opportunity to quantitatively study the scientific contributions and recognitions of scientific elites. Despite the clear difference between the Nobel laureates and "ordinary" scientists, we find universal career patterns that are applicable to both ordinary and elite scientists. Indeed, we find the careers of the laureates before winning the prize are governed by remarkably similar patterns as ordinary scientists, characterized by hot streaks and increasing reliance on team work. Hence these results help advance the canonical innovation literature by offering new empirical evidence from large-scale datasets. At the same time, we also uncovered notable but previously unknown variations along their careers associated with the Nobel prize, including shifting coauthorship structure in the prize-winning work, and a temporary but significant dip in the impact of work they produce after winning the Nobel. Overall, these results represent new empirical patterns that further enrich our understanding of careers of the scientific elite.

Together, this paper takes an initial but crucial step probing our quantitative understanding of career patterns of scientific elite, which not only offer an empirical basis for future studies of individual careers and creativity in broader domains[16,51], but also deepen our quantitative understanding of patterns governing exceptional careers in science.

## Methods

**Rescaled number of citations.** To approximate the scientific impact of each paper, we calculate the number of citations the paper received after 10 years, $C_{10}$, and use it as a proxy for the paper's impact. Previous studies[29,37,44] have shown that the average number of citations per paper changes over time. To be able to compare the impact of papers published at different times and to adjust for temporal effects, the rescaled number of citations a paper receives after 10 years, $\hat{C}_{10,i}$, is suggested as a good proxy for publication impact. According to the reference[29], given a paper i, $\hat{C}_{10,i}$, is defined as follows:

$$\hat{C}_{10,i} = 10 \cdot \frac{C_{10,i}}{\langle C_{10} \rangle}, \tag{1}$$

where $C_{10,i}$ is the raw number of 10-year citations for paper i, and $\langle C_{10} \rangle$ is the average $C_{10}$ calculated over all publications published in the same year and field.

**Definition of hit paper rate.** In Fig. 1b, we compare the "hit" paper rate—defined as the probability of publishing papers in the top 1% of rescaled 10-year citations in the same year and field—for Nobel laureates and typical authors. Our collected Nobel laureate dataset is based on information provided by the MAG, which assigns the field of subject for each paper. It is worth noting that the field of subject is a hierarchal structure with six levels. The first level contains 19 main fields, such as "physics," "chemistry," "medicine," and "biology." The second level contains 295 subfields, such as "astrophysics," "biophysics," and "geophysics." In this paper, we choose the second level fields in calculating the hit paper rate for Nobel laureates. The GS typical scientist dataset is based on information from the WOS, and it is almost impossible to precisely match the career histories of 3540 GS scientists from the WOS to the MAG. Thus, the hit rate analysis of the GS scientists is based on the WOS database itself. Papers in the WOS are also assigned to one of 234 specific field categories, such as "astronomy & astrophysics," "biophysics," and "geochemistry & geophysics." The hit paper rate for typical scientists is calculated using these 234 specific fields from the WOS.

**Selecting matching papers.** In Fig. 3b, we created a matching sample for each paper published by Nobel laureates. The procedure for selecting matching papers is introduced here in detail. For each Nobel prize-winner's work, we first determine its year of publication, total citation number, and subject categories based on the MAG dataset. Next, all the MAG papers with the same publishing year and specific field are obtained and sorted according to their number of citations. It is worth noting that when a laureate's paper spans multiple subjects, we deem MAG papers appropriate matches if they share at least one common subject with the laureate's. We then select the 20 papers with citation counts that are most similar to the laureate's paper and use these as matching papers.

**Quantifying impact.** In Fig. 4a, we compare the average impact of papers published by the laureates in each of the four years before and after winning the Nobel. We propose a measure to quantify average impact of papers: We first calculate the average impact within all papers in specific years, and then we take the individual heterogeneity of Nobel laureates into consideration when quantifying average impact of papers.

The impact of paper i, is quantified by $\Gamma_i = \log(\hat{C}_{10,i} + 1)$ where $\hat{C}_{10,i}$ measures the rescaled number of citations within 10 years of publication. We denote $\Delta y = y_i - y_{Prize}$ as a laureate's relatively publishing time after winning the Nobel Prize, where $y_i$ is the publication year of paper i. Assuming there are $N_{\Delta y}$ papers publishing in the $\Delta y$ year after winning the Prize, we define the average impact of papers as follows:

$$\langle \Gamma_P \rangle_{\Delta y} = \frac{\sum_{i=1}^{N_{\Delta y}} \Gamma_i}{N_{\Delta y}}. \tag{2}$$

However, the above measure did not consider the individual heterogeneity of Nobel laureates. For example, average impact may be driven by those laureates with high productivity as well as high paper quality. Thus, we first measure the average impact of papers for each laureate and then calculate the average for all Nobel laureates. For laureate j, the average impact of papers published in the $\Delta y$ year after winning the Prize is defined as:

$$\langle \Gamma_P \rangle_{\Delta y, j} = \frac{\sum_{i=1}^{N_{\Delta y, j}} \Gamma_i}{N_{\Delta y, j}}, \tag{3}$$

where $N_{\Delta y, j}$ is the number of papers published in the $\Delta y$ year of laureate j. Factoring in individual heterogeneity, the average impact of papers is defined as follows:

$$\langle \Gamma_N \rangle_{\Delta y} = \frac{\sum_{j=1}^{M_{\Delta y}} \langle \Gamma_P \rangle_{\Delta y, j}}{M_{\Delta y}}, \tag{4}$$

where $M_{\Delta y}$ denotes the number of laureates who still publish papers in the $\Delta y$ year after winning the Nobel Prize. In the main text (Fig. 4a), we use $\langle \Gamma_N \rangle$ to measure the average impact of papers.

**Topic changing after winning the Nobel Prize.** To quantify the topic of a paper, we adopt a recent method based on community structure of the co-citing network of a scientist's papers[51]. To ensure meaningful community detection results, we consider all Nobel laureates who have published at least 50 papers. We also excluded Nobel laureates who published fewer than five papers after winning the Prize. Finally, we selected 283 Nobel laureates (74 for physics, 96 for Chemistry, 113 for medicine) who satisfied these requirements.

In Fig. 4g, we measure the probability of two papers belonging to the same topic within 4 years before and after the reception of the Prize and a random year. To measure the probability of changing topics of Nobel laureates after winning Prize, we randomly selected two papers, within 4 years before and after the Nobel, respectively, and measured the probability of two papers belonging to the same topic. We then build a null model by randomly choosing a year as the pretended prize-winning year for comparison.

To test if Nobel laureates tend to study a new topic after winning Prize, we measure the chance of Nobel laureates shifting to a new topic after winning the Nobel, $\frac{\text{\#new topics after winning Prize}}{\text{\#topics}}$. We also shuffled the topic of the works and repeated the measurement as a null model for comparison.

**Author Contributions** D.W. and S.F. conceived the project, D. W. designed the experiments; J.L. and Y.Y. collected data and performed empirical analyses with help from S.F. and D.W.; all authors discussed and interpreted results; D.W., J.L. and Y.Y. wrote the manuscript; all authors edited the manuscript.

**Acknowledgements** The authors thank L. Liu, Y. Wang, Y. Ma, and all members of Northwestern Institute on Complex Systems (NICO) for invaluable comments. Funding data sourced from Dimensions, an inter-linked research information system provided by Digital Science (https://www.dimensions.ai).

**Data Accessibility** The main data that support the findings of this study is freely available. Deposited in public repositories with detailed descriptions in Harvard Dataverse (https://doi.org/10.7910/DVN/6NJ5RN).

**Funding Statement** This work is supported by the Air Force Office of Scientific Research under award number FA9550-15-1-0162, FA9550-17-1-0089 and FA9550-19-1-0354, National Science Foundation grant SBE 1829344 and Northwestern University's Data Science Initiative.

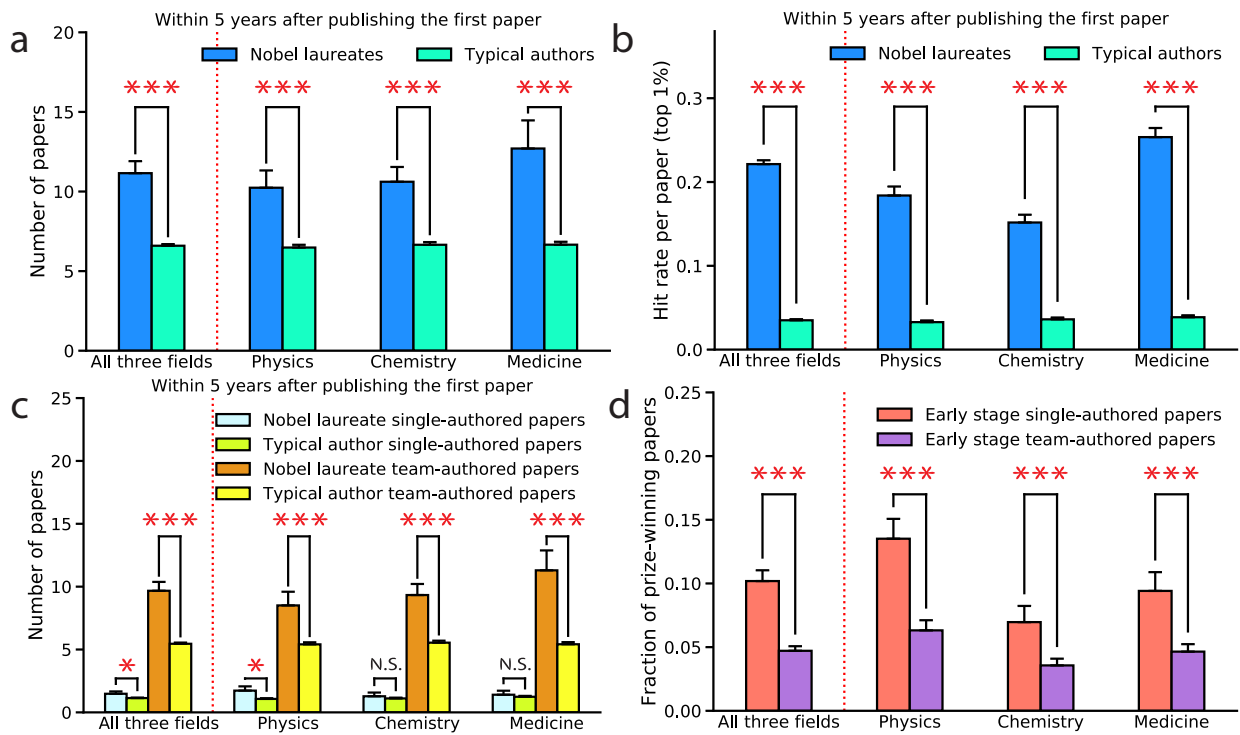

**Fig. 1 | Early Career Performance.** By early career (stage) here we mean the first five years after publishing the first paper. **a** and **b** show the early role performance of Nobel laureates compared with typical authors in terms of productivity and hit rate per paper (top 1%). We choose typical authors with at least 10 years of career length. Since most authors of the GS dataset fall in recent decades, we only consider Nobel laureates and GS typical scientists with their first paper published after 1960. We randomly select 20 typical authors with the same first-paper publishing year and research domain, eventually leading to 3540 scientists for 177 Nobel laureates. **a** In terms of productivity, the Nobel laureates are indeed more productive (11.15 vs 6.59, Student's t-test, p-value<$10^{-7}$). **b** When it comes to impact, the two populations are not comparable, and the hit paper rate (probability of publishing papers in the top 1% of rescaled 10-year citations in the same year and field) of the Nobel laureates is 6.33 times higher than for typical authors. **c** Much of the difference in early productivity between Nobel laureates and typical scientists resulted from joint papers (9.67 vs 5.46, Student's t-test, p-value<$10^{-7}$). In chemistry and medicine, there was no significant difference between the average number of single-authored papers published by laureates in their early stage and the average author. In physics, instead, Nobel laureates publish a little more single-authored papers than typical authors (Student's t-test, p-value=0.07). **d** We further compare the fractions of prize-winning papers within all laureates' early stage single-authored papers and team-authored papers published in early stages. The former is 2.16 times as high as the latter on average (Chi-squared test, p-value<$10^{-11}$). As for different disciplines, the ratios are 2.13, 1.95 and 2.03 times for physics, chemistry and medicine, respectively. ***

p<0.01, ** p<0.05, * p<0.1 and N.S. (not significant) for p>0.1. Error bars represent the standard errors of the mean.

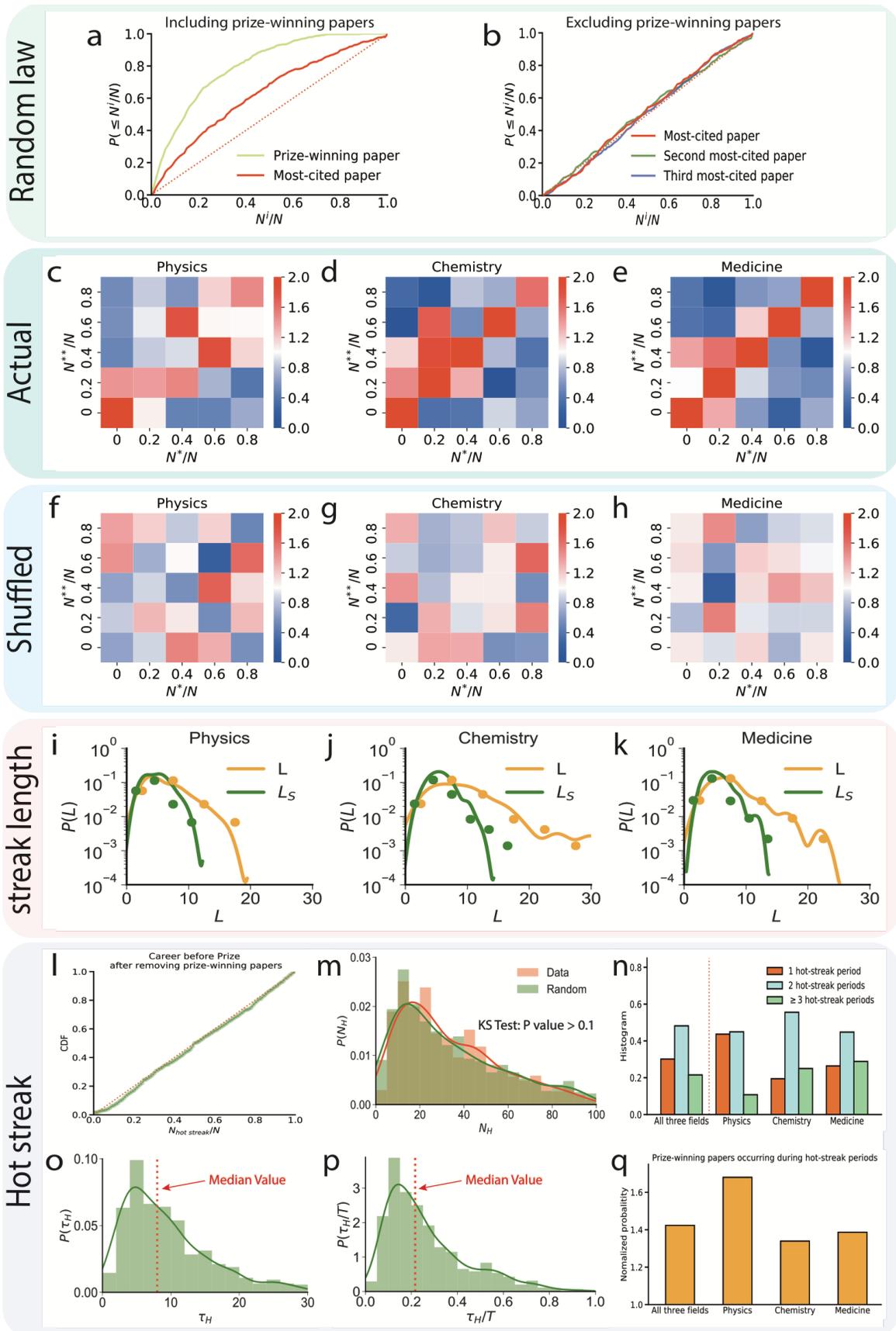

**Fig. 2 | Hot streak phenomenon. a** The cumulative distribution function P($\leq N^i$/N) of relative sequence positions of the prize-winning paper and the most-cited paper (citations are ranked based on 10-year citation counts) during the academic career before the reception of the prize. $N^i$ denotes the order of the hit work within $N$ works in a career. The red dotted line represents the null model, in which the most-cited paper can occur at any position in the sequence of papers. **b** To eliminate sample bias brought from prize-winning work (49.7% of the most-cited papers before the Prize is given are the prize-winning papers, and the Laureates wait an average of 17.6 years for formal recognition after making prize-winning achievements), prize-winning papers are removed and then we recalculate the top three most-cited papers among the papers published before the conferment of the award. **c-e** The normalized joint distribution of the relative position of the top two most cited papers (e.g., $N^*$ and $N^{**}$) within $N$ works in a career of a Nobel laureate across three domains, compared with a null model in which the two papers each follow their independent timing distributions. Values greater than one indicate that two hits are more likely to co-locate than random. **f-h** We shuffle the order of each work in a career while keeping their impact intact as a null model for **c-e**. The longest streak within a career before the Nobel Prize, L, is defined as the maximum number of consecutive works whose impact is above the median impact of the career before the Prize. **i-k** The distribution P($L$) of the longest streak within a career before the Prize and the corresponding distribution P($L_s$) for shuffled careers, for physics, chemistry and medicine, respectively. Orange dots represent empirical observations, whereas green dots correspond to shuffled careers. The orange solid line shows the simulation results produced by a hot-streak model (Supplementary Information S4.3-4.5) and the shuffled version is illustrated by the green solid line. **l** The hot-streak model describes well the laureates' scientific career pattern for different disciplines. $N_{hot\ streak}$/$N$ measures the relative position of the work lying in the middle position of the hot streak period, among works in a career before the Prize after removing the prize-winning papers. Their cumulative distributions are shown in green dots. **m** The distribution of the number of works produced during hot streaks P($N_H$), compared with a null distribution, where we randomly pick one work as the start of the hot streak for Nobel laureates. We use the Kolmogorov-Smirnov (KS) measure to compare P($N_H$) of data with the null distribution, finding that we cannot reject the hypothesis that the two distributions are drawn from the same distribution (*p*-value=0.18). **n** The histogram of the number of hot-streak periods. Nobel winners have 1.93 hot-streak periods on average, specifically, 1.67 for physics, 2.08 for chemistry and 2.04 for medicine. **o** The duration distribution of the hot streak P($\tau_H$) for Nobel laureates. The median hot streak duration $\tau_H$ is 8 years, which is shown as the red dot line. **p** The relative hot streak duration distribution P($\tau_H/T$) for Nobel laureates, where T is the career length of Nobel laureates. Red dot line shows the median relative duration. **q** We show the normalized probability of prize-winning papers

occurring during the hot-streak periods $\frac{P_{winning\ papers}}{P_{random}}$. We find that the prize-winning papers are about 1.42 times more likely occurring during the hot-streak periods than random, especially for physics laureates.

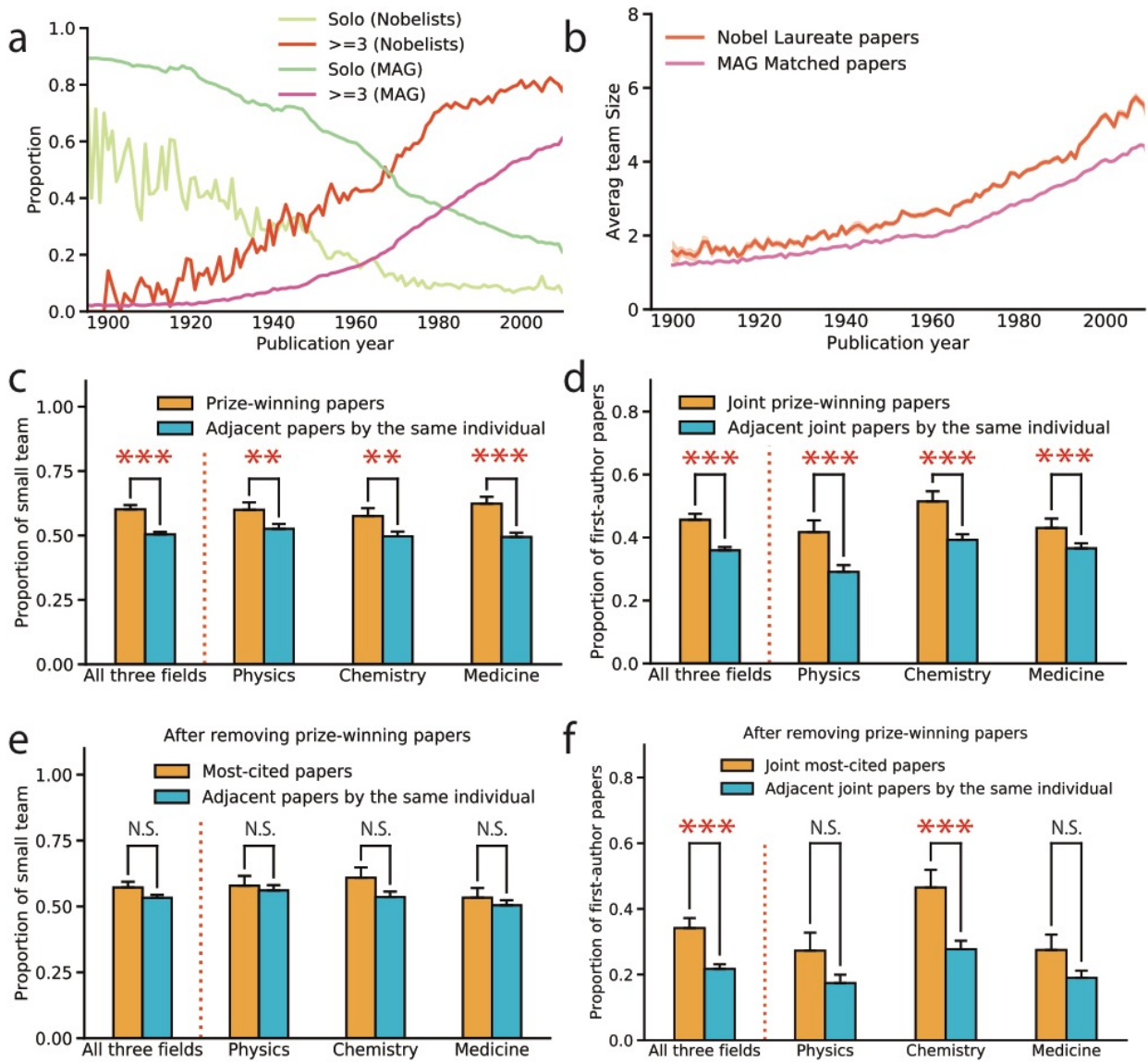

**Fig. 3 | Collaboration patterns. a** A comparison of the proportion of single-authored paper and large-team-authored paper between Nobelists' papers and all MAG papers as a function of the publishing year. The last century has seen a decline of the fraction of single-authored papers and an increase of large team collaboration. Moreover, Nobel laureates tend to have a smaller fraction of single-authored papers and a larger fraction of big team papers. **b** Team size of Nobel laureates' papers and matched papers as a function of the publication year. For each Nobel laureates' paper, we matched 20 papers with the same publication year, same specific field and the closest number of citations for comparison. It shows that the team size of Nobel laureates' papers is always larger than random. **c** The proportion of small team-sized paper (team size <=2) for all the prize-winning papers and the null model. For each prize-winning paper, we choose 4 non-prize-winning papers with the closest publication time by the same individual as a null model. **d** The proportion of first-authored paper of joint prize-winning papers in comparison with a null model. In this

case, we only consider joint-authored papers. For each joint-authored prize-winning paper, we choose 4 joint-authored non-prize-winning papers published by the same Nobel laureate with the closest publication time. **e** We also measure the proportion of small team-sized paper for the most-cited papers after removing the prize-winning papers before the Prize in comparison with a null model. For each most-cited paper, we choose 4 papers published by the same Nobel laureate with the closest publication time as the null model. **f** We then compare the proportion of first-authored papers among joint most-cited papers after removing the prize-winning papers before the Prize with a null model, and we choose 4 joint-authored papers published by the same Nobel laureate with the closest publication time as the null model. *** $p<0.01$, ** $p<0.05$, * $p<0.1$ and N.S. (not significant) for $p>0.1$. Error bars represent the standard errors of the mean.

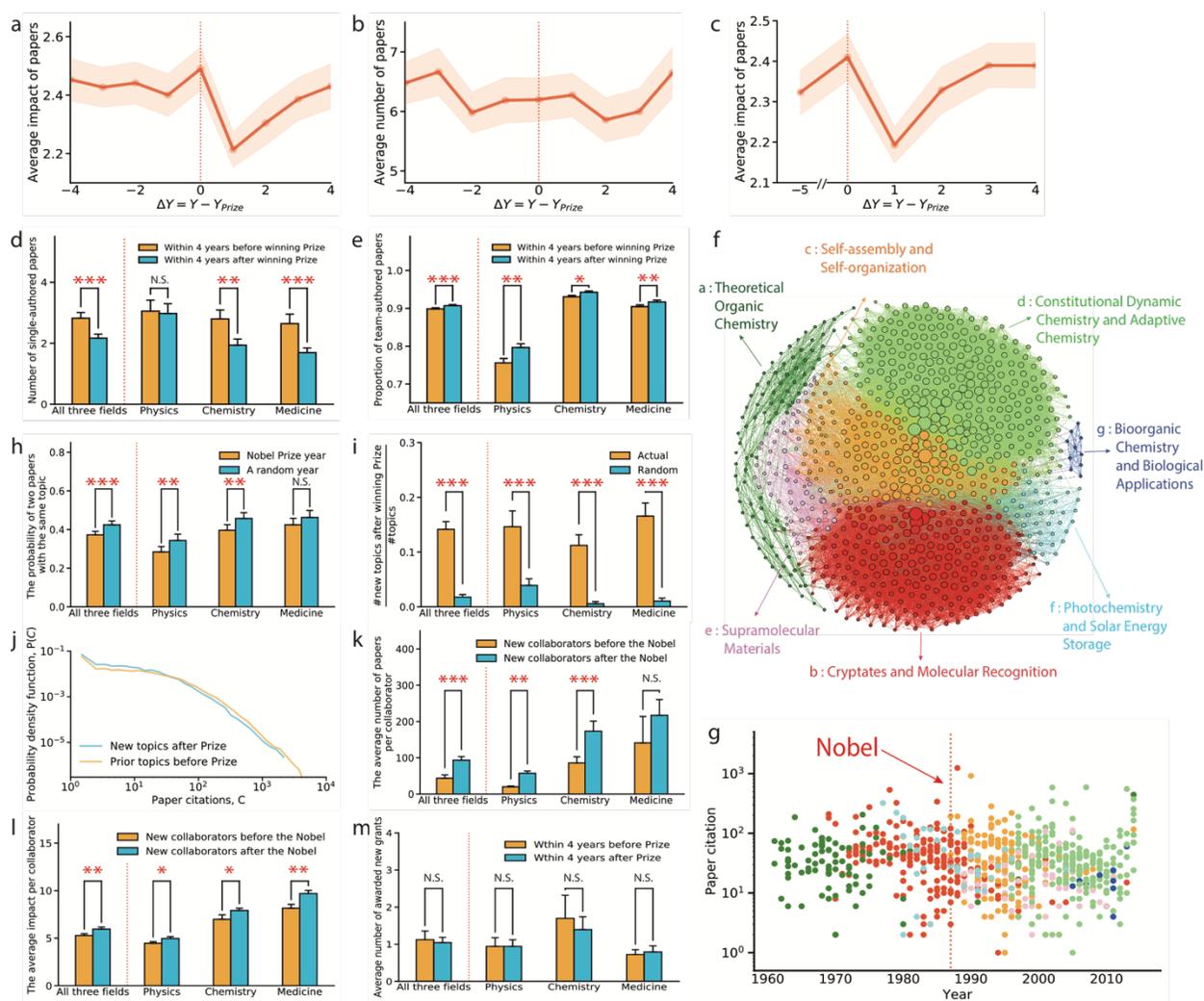

**Fig. 4 | After the Prize: The temporary dip and bounce back. a** Average impact per paper (defined in Eq. (4)) as a function of time. The year when the Prize is given is marked as 0. For each laureate, we calculate the average impact of papers she published in each of the four years before and after the Nobel, as well as the prize-winning year. The solid line indicates the average across all laureates in our sample, with the shaded area denoting the standard error of the mean. **b** The average number of papers before and after winning the Nobel. The solid line indicates the average across all laureates in our sample, with the shaded area denoting the standard error of the mean. The change following the Nobel mainly pertains to impact rather than productivity. In contrast to common belief of decreased productivity following the Nobel, the average number of publications by the laureates shows no significant changes. **c** Average impact per paper as a function of time. We set the observation window as 5 years and calculate the average impact of papers based on the 5-year citation counts. For each laureate, we compare the average impact of papers published in the 5th year before and each of the four years after winning the Nobel. **d** Comparison of the number of individual papers within 4 years before and after the reception of the Prize. It shows a significant

decrease of individual work after the Nobel. The change is significant for chemistry and medicine, while there is no significant difference for physics. **e** Comparison of the proportion of joint papers within 4 years before and after the Prize. It shows an increase of joint works after the Nobel. **f** Communities of topics for Nobel laureate Jean-Marie Lehn. Each paper is represented by a node, and two papers are connected if they share at least one reference, thus, constructing a co-citing network. Here communities are detected using the fast unfolding algorithm[43,59], and each community represents a research topic. **g** The time series of all the papers published by Nobel laureate Jean-Marie Lehn, and the Y-axis shows the citation for each paper. Each paper is represented by a point and the color corresponds to the topic community in the co-citing network. It shows a clear topic switching from 'Cryptates and Molecular Recognition' to 'Self-assembly and Self-organization' for Nobel laureate Jean-Marie Lehn right after winning Nobel prize. **h** Comparison of the probability of two papers belonging to the same topic within 4 years before and after the reception of the Prize and a random year. The probability is significant lower after winning the Prize, suggesting Nobel laureates tend to shift research topic after winning the Nobel. **i** We measure the chance of Nobel laureates shifting to a new topic after winning the Nobel, $\frac{\#new\ topics\ after\ winning\ Prize}{\#topics}$. We also shuffled the topic of the works and repeated the measurement as a null model, finding that the laureates are much more likely to shift to a new topic after winning Prize than random (14.2% vs 1.8%, $p$-value<$10^{-14}$). The change is significant for physics and chemistry, while there is no significant difference for medicine. **j** The distribution of paper citations for prior topics before Prize and new topics after Prize. We use the Kolmogorov-Smirnov measure to compare the two distribution, finding that papers of the prior topics receive higher citations than those of the new topics after Prize (Kolmogorov-Smirnov test, $p$-value<$10^{-71}$). **k** Comparison of the average number of papers of new collaborators before and after the Nobel, showing post-Prize collaborators tend to be more productive (Student's t-test, $p$-value<$10^{-3}$). **l** Comparison of the average impact of new collaborators before and after the Nobel, showing post-Prize collaborators tend to have a higher impact (Student's t-test, $p$-value=0.02)**. m** Comparison of the average number of awarded new grants within four years before and after winning the Nobel, showing there is no significant change in funding before and after the Prize (Student's t-test, $p$-value=0.77). *** $p<0.01$, ** $p<0.05$, * $p<0.1$ and N.S. (not significant) for $p>0.1$. Error bars represent the standard errors of the mean.

# Supplementary Information for Scientific elite revisited: Patterns of productivity, collaboration, authorship and impact

Jichao Li[1,2,3,4], Yian Yin[2,3,5], Santo Fortunato[6,7] and Dashun Wang[2,3,4,5*]

[1] *College of Systems Engineering, National University of Defense Technology, Changsha, China*

[2] *Center for Science of Science and Innovation, Northwestern University, Evanston, IL, USA.*

[3] *Northwestern Institute on Complex Systems, Northwestern University, Evanston, IL, USA.*

[4] *Kellogg School of Management, Northwestern University, Evanston, IL, USA*

[5] *McCormick School of Engineering, Northwestern University, Evanston, IL, USA*

[6] *School of Informatics, Computing, and Engineering, Indiana University, Bloomington, IN, USA*

[7] *Indiana University Network Science Institute (IUNI), Indiana University, Bloomington, IN, USA*

# Contents

# S1     Data Description

## S1.1 Dataset of career histories for Nobel laureates

In this project, we manually gather 874 prize-winning papers for 545 Nobel laureates. The prize-winning papers are then matched in the Microsoft Academic Graph (MAG) database. We also construct a unique dataset of publication lists of 93,394 laureate papers for almost all (545 out of 590, or 92.4%) Nobel laureates by manually combining information collected from Nobel Prize official websites, the laureates' university websites, and Wikipedia, and matching laureates with their publications from the MAG database. The dataset obtained through this process is validated with external sources from manually collected CVs, selected Google Scholar (GS) profiles, additional affiliation information, and manual verification of a random selection of 60 Nobel laureates (20 for each Nobel laureate). The collected publication records can be downloaded at https://doi.org/10.7910/DVN/6NJ5RN. The manual curation and algorithmic disambiguation procedures for collecting the publication records of the Nobel laureates are detailed in our paper[1].

## S1.2 Comparison dataset

We also leverage a dataset of over 20,040 carefully disambiguated typical authors from GS[2] as our comparison group. To show the cross-domain difference between laureates and typical authors, typical authors are classified in each of the three Nobel fields if they have published at least one paper in a related field. The Nobel field categories for physics, chemistry, and medicine are referred to Jones's paper[3]. Most typical authors in the GS dataset have produced work in recent decades. To eliminate this temporary effect, we only consider Nobel laureates and typical scientists whose first paper was published after 1960 and whose careers were at least 10 years in length. For each Nobel laureate who published her first paper after 1960, we randomly match 20 typical authors in the same research domain with the same first-paper publishing year, which eventually resulted in 3540 authors (1260 in physics, 1140 in chemistry, and 1140 in medicine) randomly matched to 177 Nobel laureates (63 in physics, 57 in chemistry, and 57 in medicine).

## S1.3 Data preprocessing

To ensure the feasibility and reliability of the results, the raw data is processed in the following ways for further analysis: 1) We only focus on the laureates' original papers, meaning that all review and note papers are removed; 2) Nobel lecture papers are also removed; 3) When performing citation analysis, we only focus on papers published before 2005 to ensure that all analyzed papers garner 10 years of citations.

# S2 Literature Review of Nobel Prize Studies

The Nobel Prize, commonly perceived as the highest honor that a natural scientist could ever have, has been frequently used as a proxy for identifying top researchers of supreme achievement. A large body of literature in disciplines ranging from sociology, economics, psychology, and physics has focused on these scientific ultra-elites, which can be summarized into two main streams: quantitative and qualitative study. On the one hand, quantitative study is based on the laureate's publication and citation records, primarily their milestone events, i.e. their prize-winning research. Topics related to these milestones include scientific creativity, credit allocation, breakthrough recognition, collaboration patterns, and changing citation dynamics. However, most of the literature is focused on either one specific domain or a small percent of the laureates, resulting from an incomprehensive dataset. On the other hand, qualitative research aims to understand laureates' entire careers, often by examining Nobel archives or interviewing prizewinners. The most representative work in this area is from Zuckerman, who interviewed 45 American Nobel laureates, systematically examining each career in detail. Table S1 reviews selected studies related to the Nobel Prize.

Table S1: Literature review of Nobel Prize studies

| Year | Paper | Category | Topic | Data |
|------|-------|----------|-------|------|
| 1957 | Manniche et al. (1957)[4] | quantitative & qualitative | Age and creativity | 164 prizewinners in the period 1901-1950 |
| 1967 | Zuckerman (1967)[5] | quantitative | Nobel prestige | Affiliation information of 76 prize-winners |
| 1967 | Zuckerman (1967)[6] | quantitative & qualitative | Productivity, collaboration and authorship | Interview information from 41 prize-winners |
| 1977 | Zuckerman (1977)[7] | quantitative & qualitative | Productivity, collaboration and authorship | Interview information from 41 prize-winners |
| 1978 | Ashton et al. (1978)[8] | quantitative | Citation analysis | Citation record of Nobel prizewinner, W. N. Lipscomb |
| 1984 | Crawford (1984)[9] | qualitative | Documented study | Nobel Archives between 1901-1915 |
| 1986 | Garfield (1986)[10] | quantitative | Citation analysis | Citation records of 125 prizewinners from 1965 to 1984 |
| 1988 | Diamond (1988)[11] | quantitative | Citation analysis | Citation counts of 24 prizewinners in economics between 1969 and 1986 |
| 1991 | MacLauchlan (1991)[12] | quantitative & qualitative | Nobel nominations study | Nobel Archives between 1901 and 1937 |
| 1992 | Garfield (1992)[13] | quantitative | Citation analysis | Citation records of 37 prizewinners between 1961 and 1976 |
| 1999 | Van Dalen (1999)[14] | quantitative | Age and creativity | Productivity record of Nobel Prize laureates in economics, 1969-1997 |
| 2000 | Rodgers (2000)[15] | qualitative | How the winners are selected | / |
| 2000 | Feldman (2000)[16] | qualitative | The history of Nobel Prize | / |
| 2001 | Friedman (2001)[17] | qualitative | Politics of excellence | Nobel Archives up to 1950 |
| 2001 | Crawford (2001)[18] | qualitative | How the winners are selected | Nobel population 1901–1950 |
| 2002 | Crawford (2002)[19] | qualitative | Social history of science | Nobel Archives between 1901 and 1939 |
| 2003 | James (2003)[20] | qualitative | Academic success | / |
| 2003 | Friedman (2003)[21] | qualitative | Prize recognition | / |
| 2005 | Hirsch (2005)[22] | quantitative | Citation analysis | Physicists who earned Nobel Prizes between 1985 and 2005 |
| 2006 | Wray (2006)[23] | qualitative | Collaboration and authorship | / |

| Year | Paper | Category | Topic | Data |
|---|---|---|---|---|
| 2008 | Rablen (2008)[24] | quantitative | Age analysis | Nobel Prize winners and nominees in physics and chemistry between 1901 and 1950 |
| 2010 | Gingras (2010)[25] | quantitative | Specialty increasing | Meta-data of nominees and winners of the chemistry and physics prizes (1901–2007) |
| 2011 | Rodríguez-Navarro (2011)[26] | quantitative | Nobel prestige | Number of Nobel Prize achievements from 1989–2008 |
| 2011 | Jones et al. (2011)[3] | quantitative | Age and creativity | A dataset of 525 Nobel Prize winners |
| 2011 | Mazloumian et al. (2011)[27] | quantitative | Citation analysis around landmark paper | Citation records for 124 Nobel Prize laureates |
| 2011 | John et al. (2011)[28] | quantitative | Recognition lag | Meta-data of Nobel laureates, 1901-2009 |
| 2013 | Heinze et al. (2013)[29] | quantitative | Citation analysis | Citation records of prizewinners Gerd Binnig, Heinrich Rohrer, Harold Kroto, Richard Smalley, and Robert Curl |
| 2013 | Chan et al. (2013)[30] | quantitative | Recognition lag | 466 recipients of Nobel Prizes in physics, Chemistry, and Physiology or medicine awarded from 1901 to 2000 |
| 2014 | Azoulay et al. (2014)[31] | quantitative | Citation analysis | 16 Howard Hughes Medical Investigator prizewinners |
| 2014 | Liu et al. (2014)[32] | quantitative | Citation analysis | Citation data of six Nobel Prize-winning articles |
| 2014 | Santo et al. (2014)[33] | quantitative | Recognition lag | Meta-data of Nobel laureates before 2014 |
| 2015 | Caroline et al. (2015)[34] | quantitative | Collaboration and authorship | Meta-data of 68 Nobel Laureates in Physiology or medicine between 1969 and 2011 |
| 2016 | Hanson et al. (2016)[35] | qualitative | How the winners are selected | / |
| 2017 | Sondhi et al. (2017)[36] | qualitative | Prize-granting entities | / |
| 2017 | Seeman (2017)[37] | qualitative | Prize recognition | Nobel Prizes for organic chemistry research since 1965 |
| 2018 | Hanson et al. (2018)[38] | quantitative | Winning formula | Hundreds of nominations for the Nobel Prize in Physiology or medicine for the period 1901–66 |
| 2018 | Ferry (2018)[39] | qualitative | Road to Prize | / |
| 2019 | Weinberg et al. (2019)[40] | quantitative | Age and creativity | Citation records of Nobel laureate economists born in or before 1926 |

# S3 Supporting Information for Figure 1

## S3.1 Hit paper rate

In Fig. 1b of the main text, we compare the "hit" paper rate—defined as the probability of publishing papers in the top 1% of citations in the same year and field—for Nobel laureates and typical authors. Our collected Nobel laureate dataset is based on information provided by the MAG, which assigns the field of subject for each paper. It is worth noting that the field of subject is a hierarchal structure with six levels. The first level contains 19 main fields, such as "physics," "chemistry," "medicine," and "biology." The second level contains 295 subfields, such as "astrophysics," "biophysics," and "geophysics." In this paper, we choose the second level fields in calculating the hit paper rate for Nobel laureates. The GS typical scientist dataset is based on information from the WOS, and it is almost impossible to precisely match the career histories of 3540 GS scientists from the WOS to the MAG. Thus, the hit rate analysis of the GS scientists is based on the WOS database itself. Papers in the WOS are also assigned to one of 234 specific field categories, such as "astronomy & astrophysics," "biophysics," and "geochemistry & geophysics." The hit paper rate for typical scientists is calculated using these 234 specific fields from the WOS.

We then compare the hit paper rate for Nobel laureates with typical authors for both single-authored and joint papers after removing prize-winning papers. As we can see from Fig. S1, the top 1% hit paper rate of single-authored papers for laureates shows a 6.7-fold increase over their peers. In the case of team-authored papers, laureates experience a 4.7-fold increase.

We also compare the top 5% hit paper rate for Nobel laureates and typical authors after removing prize-winning papers. We find that Nobel laureates experience, on average, a 3.1-fold increase over their peers. That number is 3.9-fold and 2.9-fold for single-authored paper and team-authored papers respectively (Fig. S2).

## S3.2 Robustness check of Fig. 1d

In the main text (Fig. 1d), we measure the proportion of prize-winning papers written early in a career among all single-authored papers and team-authored papers, respectively. We find that early-stage single-authored papers have a higher chance of being prize-winning than early-stage team-authored papers, especially in physics and chemistry. Does this phenomenon change over time? Here we measure the fraction of early-stage prize-winning papers as a function of year of publication (Fig. S3). Though the average fraction of early-stage prize-winning papers decreases over time, the proportion of prize-winning papers in the single-authored category is always higher than that in the team-authored category ($p$-value<$10^{-3}$). In conclusion, our results are robust over time.

We also measure the proportion of single-authored papers among all early-stage papers and prize-winning papers respectively. The fraction of single-authored papers within early-stage, prize-winning papers is much higher than that within early-stage, non-prize-winning papers (Fig. S4). Almost 40% of early-stage, prize-winning papers are single-authored, but that fraction is less than 25% of the early-stage, non-prize-winning papers category ($p$-value<$10^{-11}$). This conclusion is consistent across the three disciplines (physics: $p$-value<$10^{-5}$; Chemistry: $p$-value<$10^{-3}$; medicine: $p$-value=0.014), and the results are robust concerning different periods (Fig. S5). Although the fraction of single-authored papers is decreasing over time, the proportion of early-stage, prize-winning papers that are single-authored is still higher than that of early-stage papers more generally ($p$-value<$10^{-3}$).

More interestingly, the advantage that comes with single-authoring a paper does not remain consistent over the course of a career. Rather, early-stage scientists experience the strongest benefit from single-authoring. Such benefit decays systematically as scientist ages, reaching a balance point after roughly 30 years when prize-winning papers have relatively the same chance of being single-authored or written jointly (Fig. S6). We drew the same conclusion with respect to a person's overall odds of writing a prize-winning paper. (Fig. S7).

## S3.3 Robustness check of early career length

When we refer to "early career" or "early stage" in the main text, we mean the initial five years after a given scientist publishes her first paper. Do our findings still hold if we adjust the early career length? To answer this question, we repeated the same analysis shown in Fig. 1 for the first four and six years of the careers of both laureates and typical scientists, finding the same productivity and impact patterns (Figs. S8-9).

## S4    Supporting Information for Figure 2

### S4.1 Normalized joint probability between the three most cited papers

We quantify the relative timing between the three most cited papers ($N^*$, $N^{**}$ and $N^{***}$) within each career by calculating the joint probability $P(N^*, N^{**})$, $P(N^*, N^{***})$ and $P(N^{**}, N^{**})$ compared with a null model in which the three papers each follow their independent temporal patterns. We uncovered clear diagonal patterns across all three domains (Figs. S10a-c), showing that high-impact papers by the laureates are more likely to cluster together than expected by chance. The diagonal pattern disappeared when we shuffle the order of the works while preserving the random impact rule (Figs. S10d-f).

### S4.2 Normalized joint probability between prize-winning papers and hit works

The documented diagonal pattern prompts us to ask if a similar pattern exists between prize-winning papers and highest-impact works. Here we measure the normalized joint probability of a prize-winning paper (for Nobel laureates with multiple prize-winning papers, we take the average; the same below) and the top three highest impact works within a career before the Prize. Figures S11a-c show clear colocation diagonal patterns between prize-winning papers and the laureates' hit works. If we shuffle the order of each work in a career before the Prize while keeping each paper's impact, the pattern disappears (Figs. S11d-f).

### S4.3 Hot-streak model

We find that the career histories of laureates exhibit patterns that are broadly consistent with the hot-streak model[2]. Here, we introduce the hot-streak model. The impact of work, $i$, produced by an individual can be quantified by $\Gamma_i = \log(\widehat{C_{10}} + 1)$, where $\widehat{C_{10}}$ measures the rescaled number of citations within 10 years of publication (S3.1). Recently, a ubiquitous hot-streak phenomenon has been confirmed as a fundamental law governing a scientist's dynamics of academic performance[2]. It is shown that the vast majority of scientists have at least one hot-streak period, when an individual's performance is substantially higher than her typical performance, which can be curved with a step function:

$$\Gamma(t) = \begin{cases} \Gamma_H & t_\uparrow \leq t \leq t_\downarrow \\ \Gamma_0 & otherwise \end{cases}.$$

The function features an individual's performance deviations from a baseline performance $\Gamma_0$ at a certain point in a career $t_\uparrow$, elevating to a higher value $\Gamma_H$. After sustaining $t_H = t_\downarrow - t_\uparrow$, the individual's performance then falls back to a level similar to $\Gamma_0$.

## S4.4 Fitting the hot-streak model

The procedure to fit an individual career performance using the hot-streak model is well-documented in existing study[2]. To evaluate the goodness-of-fit of the hot-streak model, we measure the difference between fitted and real paper impact by calculating the coefficient of determination $R^2$. The coefficient of determination $R^2$ is defined with the following formula:

$$R^2 = \text{MSS}/\text{TSS} = (\text{TSS} - \text{RSS})/\text{TSS},$$

where MSS is the model sum of squares (also known as ESS, or explained sum of squares), which is the sum of the squares of the prediction minus the mean for that variable; TSS is the total sum of squares associated with the outcome variable, which is the sum of the squares of the measurements minus their mean; and RSS is the residual sum of squares, which is the sum of the squares of the measurements minus the prediction from the model.

We fit the career performance of each laureate using the hot-streak model and measure the $R^2$ between the step-function-simulated career and the real career for all Nobel winners in our dataset. We find the $R^2$ peaks around 0.75, suggesting that the hot-streak model can capture the career performance of laureates quite well (Fig. S12).

## S4.5 Parameters of the generative hot-streak model

In the main text, the solid lines of Figs. 2i-k show the simulation results of the distribution of the longest streak within a career before the Prize. In this section, we introduce how the simulation results are generated through a simple hot-streak model. Referring to the previous study[2], we use the empirical productivity information for each laureate to generate a work sequence, and the impact of a work is randomly drawn from a normal distribution $N(\Gamma_H, \sigma^2)$ if it is produced during the hot streak period, or $N(\Gamma_0, \sigma^2)$ if otherwise. We randomly pick one work out of the sequence of works if the laureate produced at least five papers before the Prize (we define a hot-streak period as being composed of no less than five papers) as the start of a hot-streak period, and denote its production year as $t_\uparrow$. We assume $\Gamma_0$, $\Gamma_H$, $\sigma$ and $t_H = t_\downarrow - t_\uparrow$ to be the same for each individual in a domain, and $\Gamma_H = \Gamma_0 + 1$ for simplicity. The value of parameters, $\Gamma_0$, $\Gamma_H$, $\sigma$, and $t_H$ of the generative hot-streak model in Figs. 2i-k are evaluated partly according to our previous fitting results and partly as a guide by eye, in particular, $\Gamma_0 = 2.9$, $\Gamma_H = 3.9$, $\sigma = 1.1$, and $t_H = 7.9$ for physics; $\Gamma_0 = 3.1$, $\Gamma_H = 4.1$, $\sigma = 0.9$, and $t_H = 9.8$ for chemistry; $\Gamma_0 = 3.0$, $\Gamma_H = 4.0$, $\sigma = 1.1$, and $t_H = 9.1$ for medicine.

### S4.6 Number and duration of hot streaks

We measure the number of hot-streak periods for laureates before the Prize (Fig. S13), and find that their careers are characterized by 1.71 hot streaks (1.51 for physics, 1.85 for chemistry, and 1.76 for medicine) on average, whereas the vast majority of scientists experience just one hot streak.

In Fig. 2o in the main text, we measure the duration distribution of the hot-streak period for Nobel laureates, finding that the duration peaks at around 5.2 years. In Figs. S15a-c, we measure the duration distribution of the hot-streak period over the course of entire careers across different disciplines. The duration peaks at around 5.0, 5.2, and 7.5 years for physics, chemistry, and medicine respectively. Dotted lines represent the median hot-streak duration, which is 8, 7, and 9 years for physics, chemistry and medicine respectively. Figs. S14a-c show the relative duration distribution for individuals, where T is the career length. Figs. S14d-f show the relative duration distribution for individuals, where T is the career length of each individual. Dotted lines represent the median relative hot-streak duration, which hovers at around 22%, 20%, and 22% for physics, chemistry and medicine respectively.

## S5 Supporting Information for Figure 3

### S5.1 Selecting adjacent papers

In Fig. 3c, we created a null model for prize-winning papers by selecting adjacent papers. The procedure for doing so is introduced here in detail. To ensure the feasibility and reliability of the results, we only focus on the laureates' original work, hence all review and note papers are removed. To rule out the temporal effect, we choose the four non-prize-winning papers with the closest publication time (within three years) for each prize-winning paper by the same individual for use as our null model. In Fig. 3d, we also created a null model for joint prize-winning papers for comparison. In that case, we only consider the joint-authored papers, and for each joint-authored prize-winning paper, we choose 4 joint-authored non-prize-winning papers published by the same Nobel laureate with the closest publication time as the null model.

The same procedure for selecting adjacent papers is also used to build a null model for the hit works (ranked by 10-year citations) to use in comparison when analyzing team size and authorship. When it comes to hit works, it is worth noting that we only focus on papers published before 2005 to ensure that all analyzed papers garner 10-year citations.

### S5.2 Authorship analysis controlling for temporal factors

To check if the authorship pattern documented in Fig. 3d is dominated by a temporal effect, we control for publication time and age information using a logistic regression model. The logistic regression model is defined as follows:

$$\log\left(\frac{p_i}{1-p_i}\right) = \beta_0 + \beta_1 x_{i1} + \beta_2 x_{i2} + \beta_3 x_{i3} + \varepsilon_i,$$

where $p_i$ indicates the probability of paper $i$ having the laureate as the first author; $x_{i1}$ is a dummy variable, $x_{i1} = 1$ if paper $i$ is a prize-winning paper, $x_{i1} = 0$ if otherwise; $x_{i2}$ is the publication year of paper $i$; and $x_{i3}$ denotes the career age of the laureate when publishing paper $i$. For each joint-authored prize-winning paper, we choose four joint-authored non-prize-winning papers published by the same Nobel laureate with the closest publication time. We find that the results are robust after controlling for temporal effect, which supports our conclusion that laureates are more likely to claim first authorship in the case of their prize-winning papers ($p$-value<$10^{-3}$).

### S5.3 Authorship analysis for hit works

We have studied the authorship of all joint prize-winning papers in Fig. 3d, finding that prize-winning papers are substantially more likely to be first-authored by laureates. To test if this phenomenon is unique to the prize-winning work, we repeated the same analysis for each laureate's top three most-cited papers before the Prize by comparing them with those published by the same laureates immediately before and after. First, we compared the proportion of first-authored papers of joint most-cited papers to a null model. For each joint-authored most-cited paper, we choose four joint-authored papers published by the same Nobel laureate with the closest publication time and used these for the null model. While we find that a laureate is more likely to have first-authored her most-cited paper, the phenomenon for the second and third most-cited papers in terms of different disciplines is pretty weak (Figs. S15a-c). It is worth noting that 51.74% of the laureate's most-cited papers are also his/her prize-winning papers. To test if this phenomenon is driven by the recognition that comes with prize-winning work, we repeated our measurements by removing the careers of laureates who published their prize-winning work, and find no statistical difference for physics and medicine, while the difference still holds for chemistry (Figs. S15d-f).

## S6 Supporting Information for Figure 4

### S6.1 Hit paper rate changes after the Prize

How does a laureate's hit paper rate change after a Prize is awarded? To answer this question, we first measure the top 1% hit paper rate as a function of career years, with the year the Prize is given marked as 0. For each laureate we calculate the top 1% hit paper rate of papers she published in each of the four years before and after the Nobel, as well as those in the prize-winning year. Then, we calculate the average top 1% hit paper rate across all laureates in our sample. We find that the average top 1% hit paper rate remains almost the same—around 17% four years before the Prize. It shows a significant drop in the subsequent two years after the Nobel is awarded, reaching 12%, and then increases slightly by year four. (Fig. S16a). We also measure changes in the average top 5% hit paper rate after the Prize is awarded. It also experiences a significant drop in the two years after the Nobel, with a slight increase by year four. (Fig. S16b).

### S6.2 Robustness check of disciplines

To study if the Nobel dip effect occurs in different disciplines, we calculate the average impact of papers published by the laureates in each of the four years before and after winning the Nobel across three domains. We find that the Nobel dip is most pronounced for prizewinners in physics, who show a reduced impact of 18.1% the second year after the Prize, compared with 4.8% for chemistry and 13.4% for medicine, where prizewinners experiences their most significant dip in the third and first year respectively (Fig. S17). In all domains, the prizewinners' impact gradually recovers, bouncing back to a pre-Nobel level by year four. However, there is no significant difference when it comes to the laureate's productivity before and after the Prize (Fig. S18).

### S6.3 Robustness check of publication time

Nobel dip rate, denoted as R in our study, is defined as follows:
$$R = \frac{\langle \Gamma_N \rangle_{after} - \langle \Gamma_N \rangle_{before}}{\langle \Gamma_N \rangle_{after}},$$
where $\langle \Gamma_N \rangle_{before}$ is the average impact of papers two years before winning the Nobel Prize, while $\langle \Gamma_N \rangle_{after}$ measures the average impact of papers two years after winning the Nobel Prize.

To check if our results hold over time, we measure the Nobel dip rate according to each decade in the last century (we merge the first 30 years of the last century into one group because of the small amount of data in the sample). The average impact of papers shows a significant drop in the two years immediately following the Nobel, and the results are robust with respect to different time periods (Fig. S19).

### S6.4 Nobel dip after controlling for the burst

The burst of citations to pre-Prize paper may account for part of the observed Nobel dip. Yet more interestingly, we find that the dip persists after we control for the burst. More specifically, we re-measure the paper impact based on citations received within an observation window $k$, i.e., $k$ years of publication. The impact of paper i, is quantified by $\Gamma_i = \log(\hat{C}_{k,i} + 1)$, where $\hat{C}_{k,i}$ measures the rescaled number of citations within $k$ years of publication. For each laureate, the average impact of papers published in the $j_{th}$ year after Prize is defined as $\langle \Gamma \rangle_j = \frac{\sum_{i=1}^{N_j} \Gamma_i}{N_j}$, where $N_j$ is the number of papers published in the $j_{th}$ year. Given an observation window $k$, we exclude all works where the impact measure includes post-Prize citations for the pre-Nobel papers and compare the impact of works published in the kth years before and each of the years after winning

Prize. In the main text, we set the observation window as 5 years, and here we alter the observation window as 4 years, 3 years and 2 years. Then we calculate the average impact per paper as a function of publication year with the different observation, we find the average impact per paper still shows a significant drop the year following the Nobel even after excluding works where the impact measure includes post-Prize citations. (Figs. S20a-c).

## S6.5 Topic changing after winning the Nobel Prize

To quantify the topic of a paper, we adopt a recent method based on community structure of the co-citing network of a scientist's papers[41]. To ensure meaningful community detection results, we consider all Nobel laureates who have published at least 50 papers. We also excluded Nobel laureates who published fewer than five papers after winning the Prize. Finally, we selected 283 Nobel laureates (74 for physics, 96 for Chemistry, 113 for medicine) who satisfied these requirements.

In the main text (Fig. 4h), we measure the probability of two papers belonging to the same topic within 4 years before and after the reception of the Prize and a random year. To measure the probability of changing topics of Nobel laureates after winning Prize, we randomly selected two papers, within 4 years before and after the Nobel, respectively, and measured the probability of two papers belonging to the same topic. We then build a null model by randomly choosing a year as the pretended prize-winning year for comparison.

To ask if Nobel laureates tend to study a new topic after winning Prize, we then measure the chance of Nobel laureates shifting to a new topic after winning the Nobel, $\frac{\#new\ topics\ after\ winning\ Prize}{\#topics}$. We also shuffled the topic of the works and repeated the measurement as a null model for comparison.

## S6.6 Topic identification based on Infomap algorithm

In the main text (Fig. 4), we find that laureates disproportionately shift research topic after winning the Nobel. To test if the results are affected by the choice of different community detecting methods, we repeated the results based on another algorithm, Infomap[42]. Figure S21 visualizes the publication history of Nobel laureate Jean-Marie Lehn by color coding each paper according to the topic community to which it belongs. Based on the Infomap community detecting algorithm, the papers of Nobel laureate Jean-Marie Lehn are divided into 17 different communities of topics with 7 starting before the Nobel and 10 starting after (including the Noble year) the Nobel.

We then measured measure the probability of changing topics of Nobel laureates after winning Prize based on the Infomap community detecting method, finding only 42.3% of the two papers belonging to the same topic after winning the Prize. We further build a null model by randomly choosing a year as the pretended prize-winning year for comparison, finding that the probability is significant higher (48.5% vs 42.3%, *p*-

value=0.03, Fig. S22), which suggests Nobel laureates tend to shift research topic after winning the Nobel.

We also measured the chance of Nobel laureates shifting to a new topic after winning the Nobel based on the Infomap community detecting method, finding the laureates have more than one quarter chance of shifting to a new topic after winning the Nobel. However, the chance is only around 5.0% when we shuffled the topic of the works and repeated the measurement as a null model (25.1% vs 5.0%, $p$-value<$10^{-12}$, Fig. S23).

### S6.7 The popularity of research topics before and after the Prize

To test if the overpopulation of prior topics might encourage laureates to change research agenda after the Nobel, we first focus on laureates' papers that belong to the topics recognized by the Nobel and use citations to these papers as a proxy to trace topic popularity. We find that papers of the prior topics receive much higher citation counts after the Prize than before (Fig. S24, Student's t-test, $p$-value = 0.04). We then compared the total citation counts of the prior topics before the Prize and those of the new topics after the Prize. We find that the former received higher citations than the latter (Kolmogorov-Smirnov test, $p$-value < $10^{-71}$, Fig. 4j in the main text).

While it's difficult to measure directly whether a topic is over-popular, these results suggest that (1) the prior topic continues to be popular after the Nobel; (2) papers published on the Nobel topic tend to garner more citations than papers on the new topic. Together these results are consistent with an endogenous shift by the laureates to explore new directions, even though the new topic may be less popular (less cited on average) than the Nobel topic.

### S6.8 Changes in collaborators before and after the Prize

Another rather insightful mechanism is that the Nobel may attract new collaborations, which may be related to the topic change observed following the Nobel. To test this hypothesis, we first compare the number of new collaborators before and after the Prize. We find there is no significant change in the number of new collaborators before and after the Prize (Fig. S25), a pattern that is consistent across all three domains.

Although the number of new collaborators did show any significant change, the Nobel with added recognition may attract more established collaborators. To test this hypothesis, we analyzed the publication record of the collaborators before and after the Nobel. For each collaborator, we only focus their career record before the first collaboration with the laureate. We define $N_k$ as the number of papers published before the first collaboration with Nobel laureates for collaborators k, and the impact of collaborators k is measured as $\Gamma_k = \log(\sum_{i=1}^{N_k} C_{10,i} + 1)$, where $C_{10,i}$ is the 10-year citation counts of paper i. Comparing the profiles of collaborators before and after the Nobel, we find post-Prize collaborators tend to be more productive (Student's t-test, $p$-value < $10^{-3}$, Fig. 4k in the main text) and have a higher impact (Student's t-test, $p$-value = 0.02, Fig. 4l in the main text).

Together, these results suggest that the Nobel does not necessarily attract more new collaborators than expected, but new collaborators after the Nobel tend to be more productive with higher impact. From this result, one would expect the impact of works produced after the Nobel to increase, not decrease, which then makes the impact dip shortly after the Nobel more intriguing.

## S6.9 Funding opportunities before and after the Prize

The Prize and the added recognition may result in more and better funding opportunities, conducive to changing research directions. To test this hypothesis, in this revision we have acquired a novel dataset that captures funding records across all funding agencies world-wide, linked with individual investigators and resulting publications[43-47]. To the best of our knowledge, it is the most comprehensive research grants database to date[48], ideal to test this hypothesis.

We collected around 5 million grants between 1950 and 2019. We then connect the grant records to the careers of Nobel laureates we curated through systematic efforts of paper matching and author name disambiguation, resulting in 470 grants for 51 Physics laureates, 920 grants for 66 Chemistry laureates, and 509 grants for 72 Medicine laureates, respectively.

To measure changes in funding opportunities before and after the Prize, we first compare the average number of active grants for each laureate in each of the four years before and after winning the Nobel, finding that overall there is no significant change (Fig. S26a). We then separated this measurement by field, finding a suggestive signal for Physics laureates to increase their grant support around the time of the award (Fig. S26b), whereas the active grants for Chemistry and Medicine laureates stayed roughly the same (Figs. S26cd). We also compare the average number of new grants within four years before and after winning the Nobel, arriving at the same conclusion (Fig. 4m in the main text). Together our analysis suggests that there is no significant change in funding before and after the Prize, which is somewhat surprising. Indeed, according to the Matthew effect, we would have expected to observe an increase in funding. On the other hand, the lack of increase is consistent with the change in topic and research agenda which requires more time to build up for funding.

# S7 Supplementary Figures

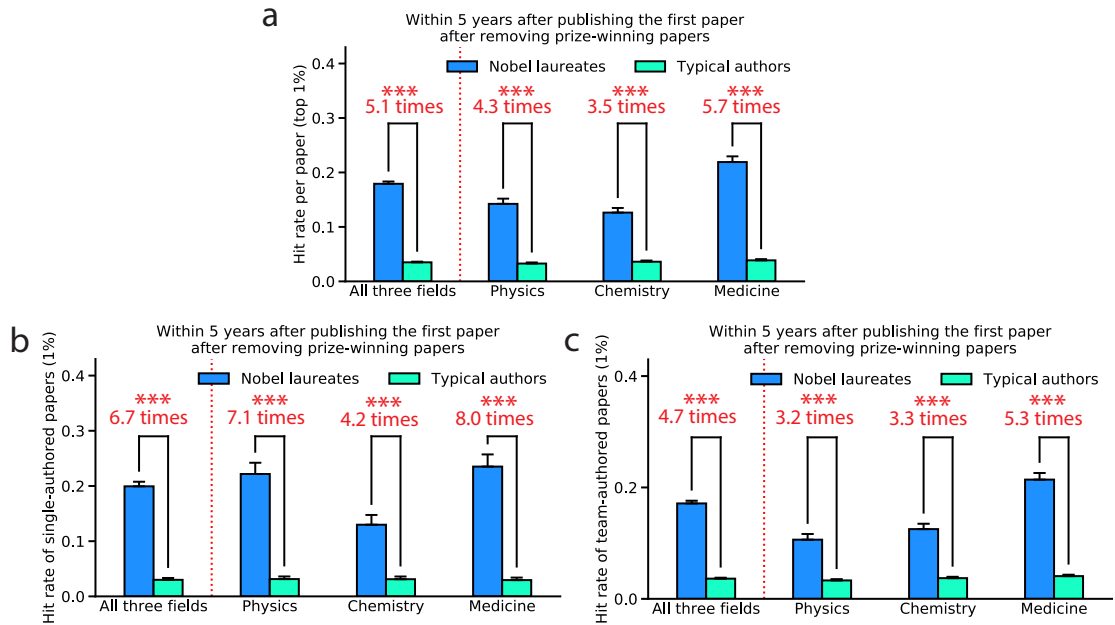

**Fig. S1 | Hit paper rate (top 1%) after removing the prize-winning paper.** A comparison of hit paper rate (top 1%) for Nobel laureates and typical authors in terms of all papers (**a**), single-authored papers (**b**), and joint papers (**c**). Even after removing prize-winning papers, both Nobel laureates' early-stage single-authored papers and joint-authored papers have much higher impact in terms of the hit paper rate (top 1%). *** $p<0.01$, ** $p<0.05$, * $p<0.1$ and NS for $p>0.1$. Error bar represents the standard error of the mean.

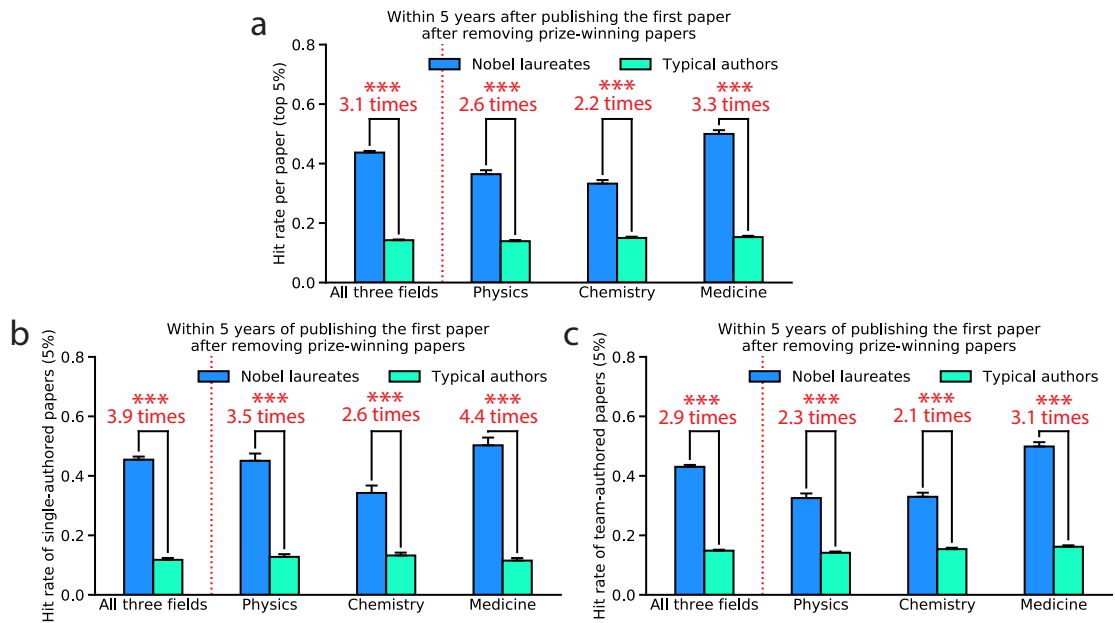

**Fig. S2 | Hit paper rate (top 5%) per early-stage paper after removing prize-winning paper.** A comparison of hit paper rate (top 5%) for Nobel laureates and typical authors in terms of all papers (**a**), single-authored papers (**b**), and joint papers (**c**). Even after removing prize-winning papers, both Nobel laureates' early-stage single-authored papers and joint-authored papers have much higher

impact in terms of the hit paper rate (top 5%). *** p<0.01, ** p<0.05, * p<0.1 and NS for p>0.1. Error bar represents the standard error of the mean.

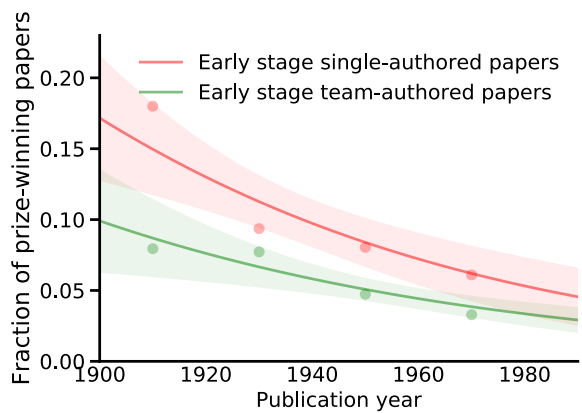

**Fig. S3 | Proportion of early-stage prize-winning papers as a function of publication time.** The average fraction of early-stage prize-winning papers decreases over time, which is well-fitted using a logistic regression model. The scattered data points refer to the mean value averaged within 20 years and shading around the lines shows 95% confidence limits. In the early-stage, after controlling for publication year, we also find that single-authored papers have a higher probability of being prize-winning than team-authored papers ($p$-value$<10^{-3}$).

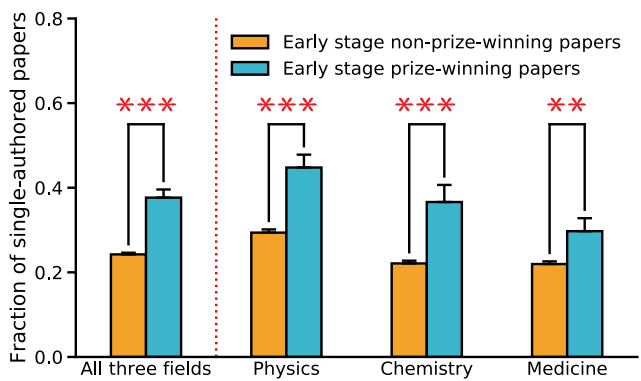

**Fig. S4 | Fraction of single-authored papers in the early stage.** We compare the fraction of single-authored papers within all laureates' early stage non-prize-winning papers (orange) and within prize-winning papers published in early stages (green) for the entire group of laureates that we examined. The fraction of single-authored papers is much higher among early-stage prize-winning papers than for non-prize-winning papers in the early stage, a trend that holds for physics, Chemistry, and medicine. *** p<0.01, ** p<0.05, * p<0.1 and N.S. (not significant) for p>0.1. Error bars represent the standard errors of the mean.

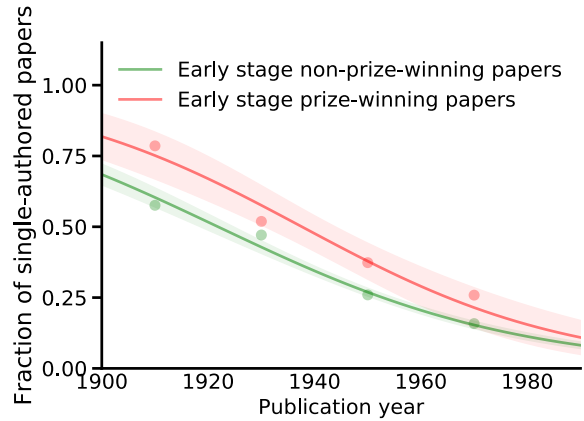

**Fig. S5 | Proportion of single-authored early stage papers as a function of publication time.** The average fraction of single-authored early stage papers decreases over time, which is well-fitted using a logistic regression model. The scattered data points refer to the mean value averaged within 20 years and shading around the lines shows 95% confidence limits. After controlling for the publication year, we find that early-stage papers are more likely to be single-authored ($p$-value<$10^{-3}$).

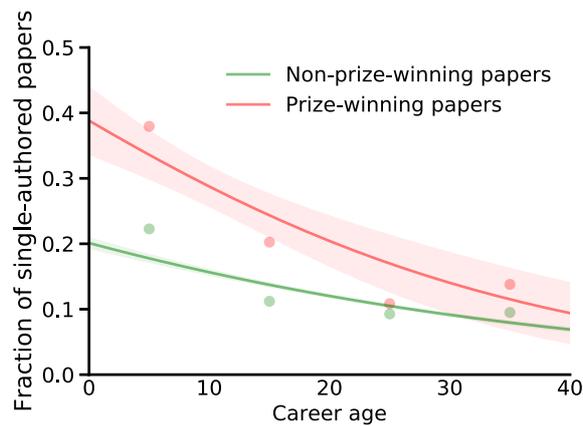

**Fig. S6 | Proportion of single-authored papers as a function of career age.** The average fraction of single-authored papers decreases over career age, which is well-fitted using a logistic regression model. The scattered data points refer to the mean value averaged within 10 career years and shading around the lines shows 95% confidence limits. The fraction of single-authored papers is much higher among prize-winning than non-prize-winning papers in the early stage, though the advantage of single-authoring decays systematically over time, reaching a balance point after roughly 30 years, when prize-winning papers have relatively the same chance of being single-authored or team-authored.

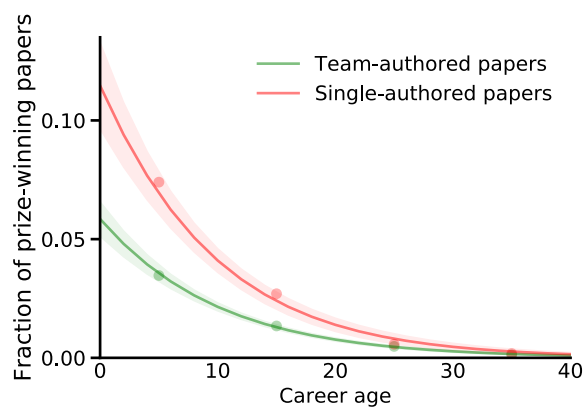

**Fig. S7 | Proportion of prize-winning papers as a function of career age.** The average fraction of prize-winning papers decreases over career age, which is well-fitted using a logistic regression model. The scattered data points refer to the mean value averaged within 10 career years and shading around the lines shows 95% confidence limits. Single-authored papers have a higher chance of being prize-winning in the early stage, though the advantage of single-authoring decays systematically over time, reaching a balance point after roughly 30 years, when single-authored and team-authored papers have relatively even odds of being prize-winning.

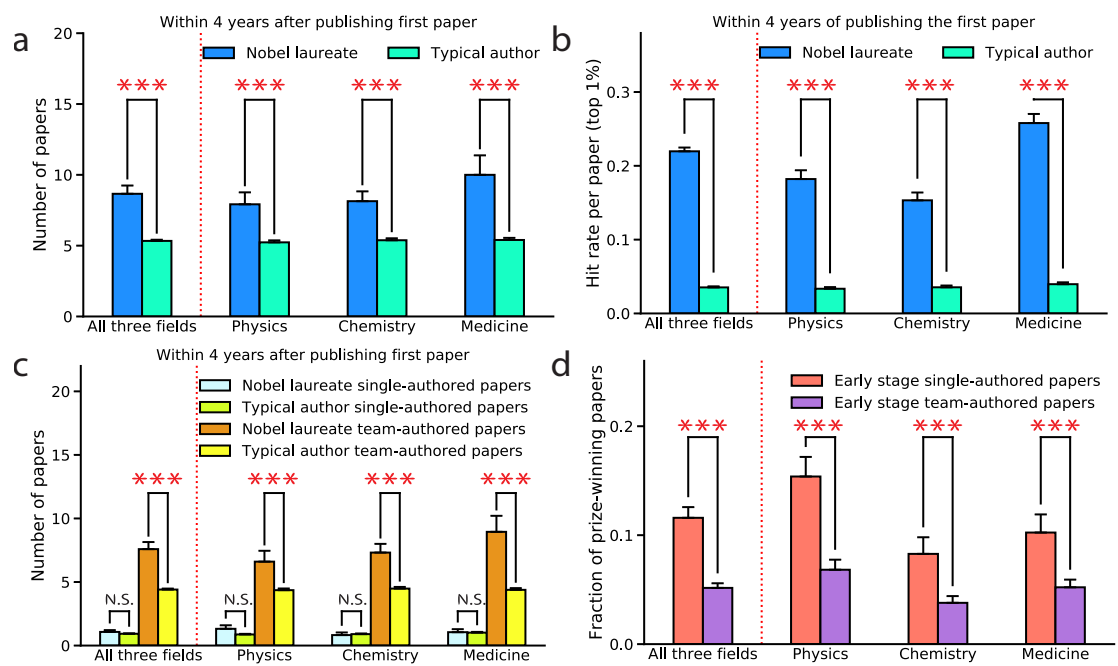

**Fig. S8 | Early Career Performance (The First Four Years). a** In terms of productivity, the Nobel laureates are indeed more prolific (8.66 vs 5.34, Student's t-test, $p$-value<$10^{-7}$). **b** When it comes to impact, the two populations are not comparable, and the hit paper rate (top 1%) of the Nobel laureates is 6.23 times higher than for typical authors. **c** Much of the difference in early published output between Nobel laureates and typical scientists resulted from joint papers (7.59 vs 4.41, Student's t-test, $p$-value<$10^{-7}$). There was no significant difference between the average number of single-authored papers published by laureates in their early stage and the average author. **d** We further compare the fractions of prize-winning papers within all laureates' early stage single-authored papers and within team-authored papers published in early stages. The fraction of prize-

winning papers is much higher among early-stage single-authored papers than for team-authored papers (Chi-squared test, $p$-value<$10^{-11}$). *** p<0.01, ** p<0.05, * p<0.1 and N.S. (not significant) for p>0.1. Error bars represent the standard errors of the mean.

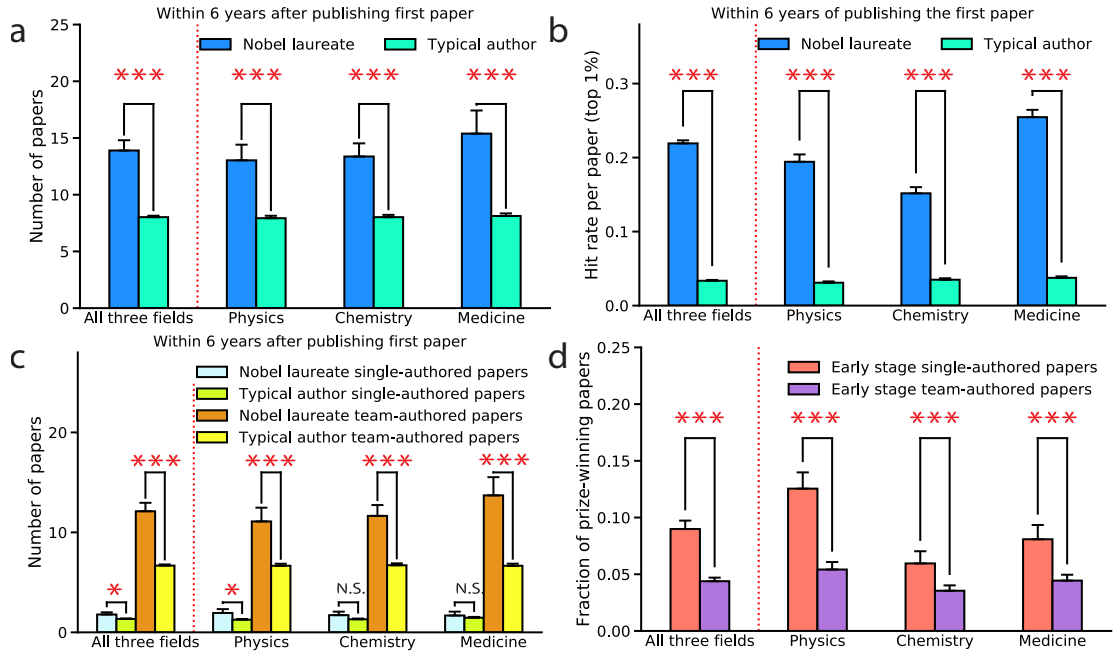

**Fig. S9 | Early Career Performance (The First Six Years). a** In terms of productivity, the Nobel laureates are indeed more prolific (13.90 vs 8.02, Student's t-test, $p$-value<$10^{-8}$). **b** When it comes to impact, the two populations are not comparable, and the hit paper rate (top 1%) of the Nobel laureates is 6.53 times higher than for typical authors. **c** Much of the difference in early published output between Nobel laureates and typical scientists resulted from joint papers (12.11 vs 6.68, Student's t-test, $p$-value<$10^{-8}$). In chemistry and medicine, there was no significant difference between the average number of single-authored papers published by laureates in their early stage and the average author. In physics, Nobel laureates tend to publish more single-authored papers than typical authors (Student's t-test, $p$-value=0.09). **d** We further compare the fractions of prize-winning papers within all laureates' early stage single-authored papers and within team-authored papers published in early stages. The fraction of prize-winning papers is much higher among early-stage single-authored papers than for team-authored papers (Chi-squared test, $p$-value<$10^{-10}$). The difference is significant for laureates in physics and Chemistry, but we find no significant difference for laureates in medicine. *** p<0.01, ** p<0.05, * p<0.1 and N.S. (not significant) for p>0.1. Error bars represent the standard errors of the mean.

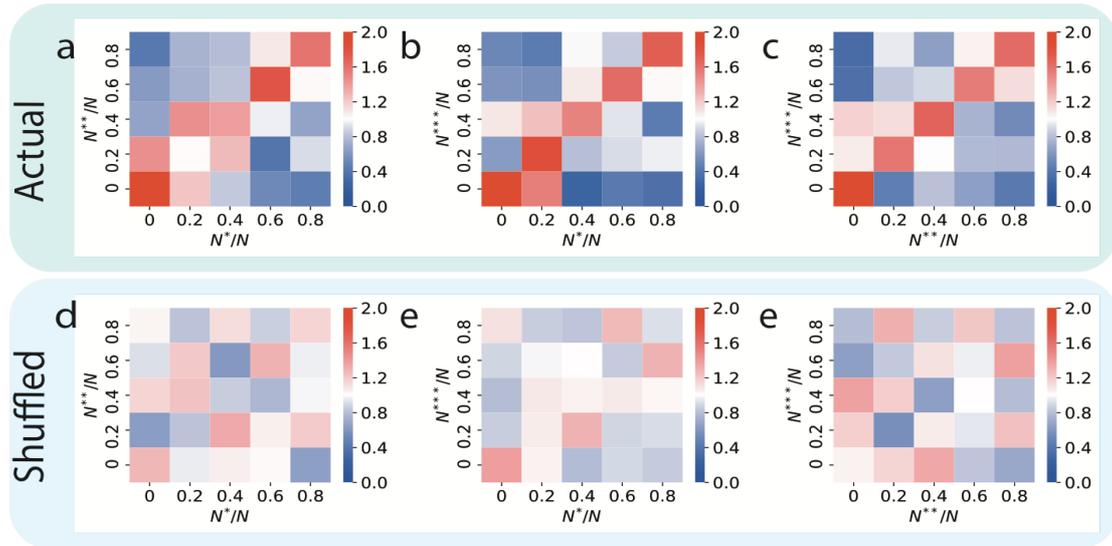

**Fig. S10 | Joint probability of the top three highest impact works. a-c** We measure the joint probability of the top three highest impact works within a career before the Prize. Values greater than 1 indicate two hits are more likely to collocate than random. **d-f** We shuffle the order of each work in a career while keeping their impacts intact as a null model for **a-c**. We find clear diagonal patterns, demonstrating that high-impact papers are more likely to cluster together than expected by chance. The diagonal pattern disappears when we shuffle the order of the works.

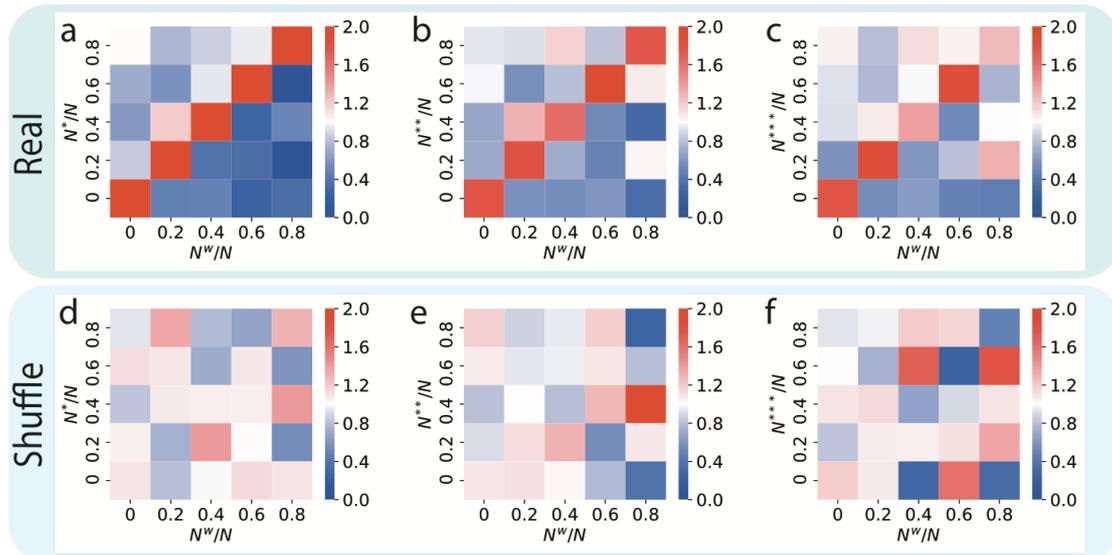

**Fig. S11 | Joint probability of prize-winning papers and the top three highest impact works. a-c** We measures the joint probability of a prize-winning paper (for Nobel laureates with multiple prize-winning papers, we take the average; the same below) and the top three highest impact works within a career before the Prize. Values greater than 1 indicate two hits are more likely to collocate than random. **d-f** We shuffle the order of each work in a career while keeping their impact intact as a null model for **a-c**. We can find a clear positive correlation between prize-winning papers and the top three highest impact works along the diagonal.

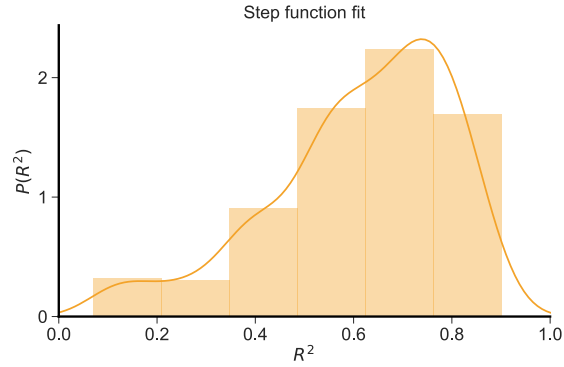

**Fig. S12 | The distribution of the coefficient of determination $R^2$.** We fit each laureate's career performance using the hot-streak model, and measure the $R^2$ between the step-function-simulated careers and the real careers of all Nobel winners in our dataset. We find the $R^2$ peaks around 0.75, suggesting the hot-streak model captures the career performance of laureates well.

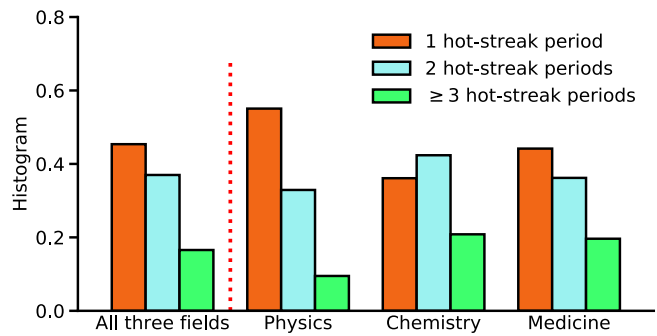

**Fig. S13 | Histogram of the number of hot-streak periods before the Prize.** Laureates in all three Nobel fields experience 1.71 hot-streak periods on average (1.51 for physics, 1.85 for chemistry and 1.76 for medicine).

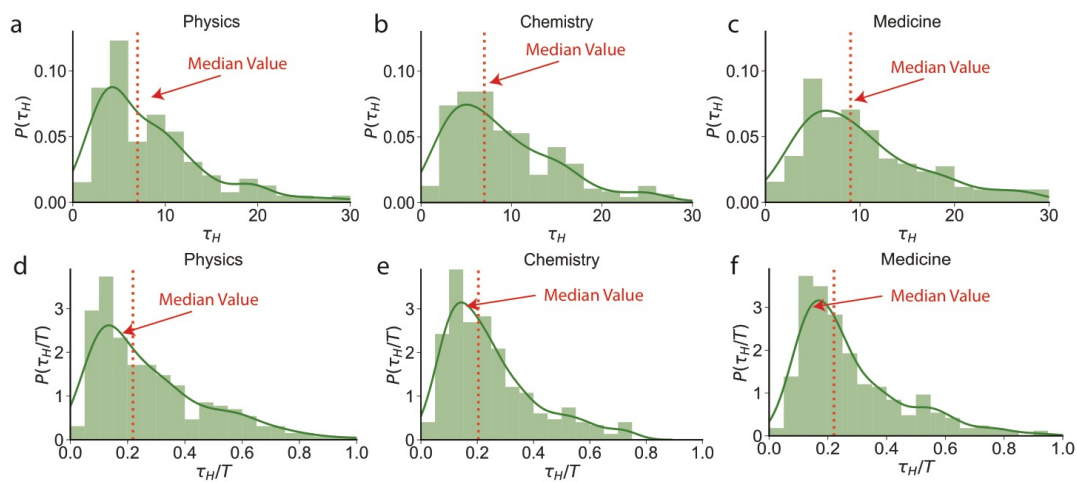

**Fig. S14 | Hot-streak duration distribution across different disciplines. a-c** We measure the duration distribution of the hot-streak period within careers across different disciplines. The duration peaks at around 5.0, 5.2 and 7.5 years for physics, chemistry and medicine respectively. Dotted lines represent the median hot-period duration, which is 8, 7, and 9 years for physics, chemistry and medicine respectively. **d-f** We shows the relative duration distribution for individuals, where T is

the career length of each individual. Dotted lines represent the median relative hot-period duration, which hovers around 22%, 20% and 22% for physics, chemistry and medicine respectively.

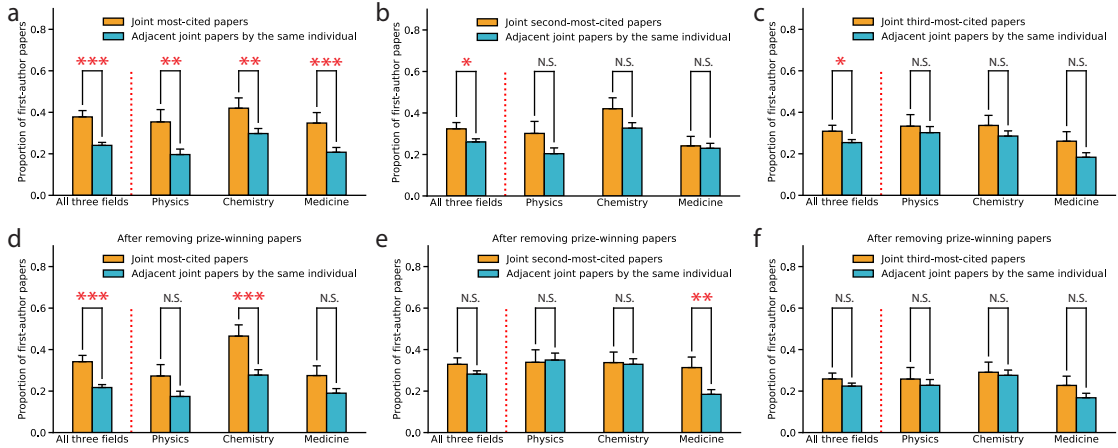

**Fig. S15 | Fraction of first-authored papers of joint hit works.** We calculate the laureates' fraction of first-authored papers of joint hit works before the Prize (**a** for most-cited papers, **b** for second-most-cited papers, and **c** for third-most-cited papers) in comparison with a null model. In this case, we only consider joint-authored papers. For each joint-authored hit work, we choose four joint-authored papers published by the same Nobel laureate with the closest publication time as the null model. To rule out the effect of the prize-winning works, we repeated our measurements by removing the careers of laurates who published their prize-winning work (**d** for most-cited papers, **e** for second-most-cited papers, and **f** for third-most-cited papers). *** $p<0.01$, ** $p<0.05$, * $p<0.1$ and N.S. (not significant) for $p>0.1$. Error bars represent the standard errors of the mean.

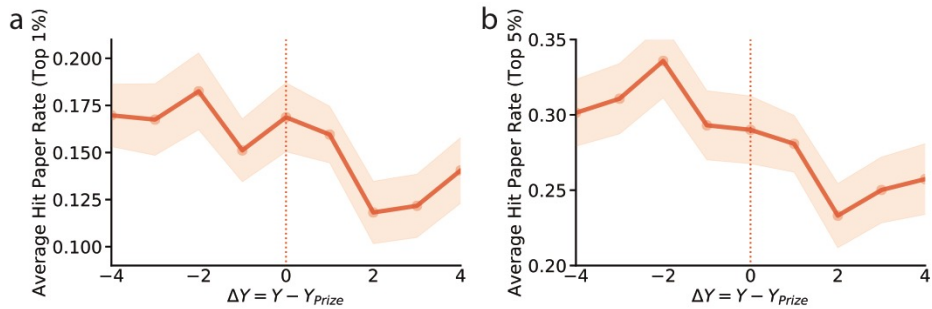

**Fig. S16 | Average hit rate of papers published four years before and after the Nobel Prize year. a** The solid line indicates the average across all papers in our sample, with the shaded area denoting the standard error of the mean. We find that the average top 1% hit paper rate remains almost the same--roughly 17%--four years before the Prize, then shows a significant drop in the subsequent two years after the Nobel is awarded, reaching 12%, and then slightly increasing by year four. **b** We also measure the average top 5% hit paper rate change before and after the Prize, which likewise experiences a significant drop in the two years after the Nobel is awarded and a slight increase by year four.

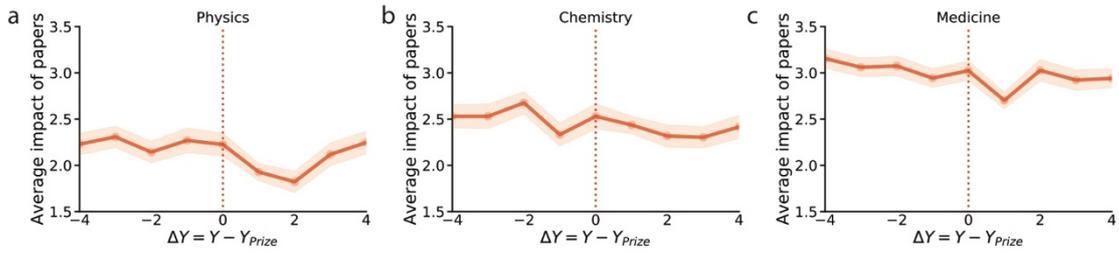

**Fig. S17 | Impact of papers published four years before and after the Nobel Prize year according to different disciplines.** The solid line indicates the average across all laureates in our sample, with the shaded area denoting the standard error of the mean. The Nobel dip is most pronounced in physics, where laureates show a reduced impact of 18.1% in the second year after the Prize, compared with a drop of 4.8% in chemistry and 13.4% in medicine. In chemistry and medicine, laureates experience their most significant dip in the third and first year respectively.

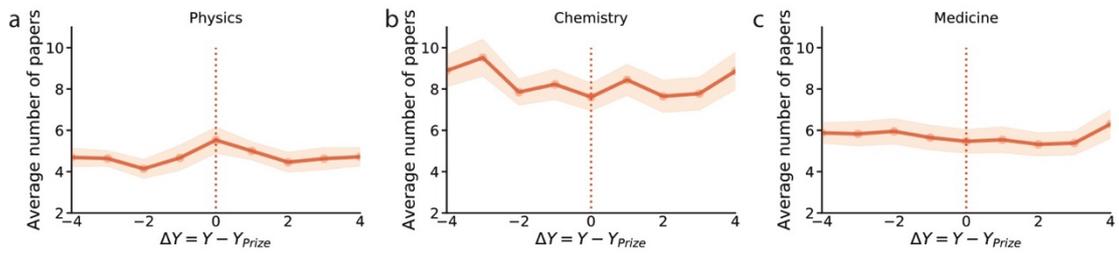

**Fig. S18 | Average number of papers published four years before and after the Nobel Prize year according to different disciplines.** The solid line indicates the average across all laureates in our sample, with the shaded area denoting the standard error of the mean. In all three disciplines, productivity does not drop significantly after the Nobel is awarded.

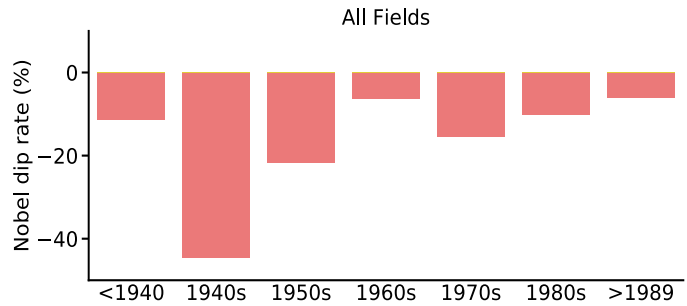

**Fig. S19 | Nobel dip rate according to different time periods (two years before and after winning Nobel Prize).** The average paper impact per paper shows a significant drop in the two years immediately following the Nobel, and these results are robust with respect to different time periods.

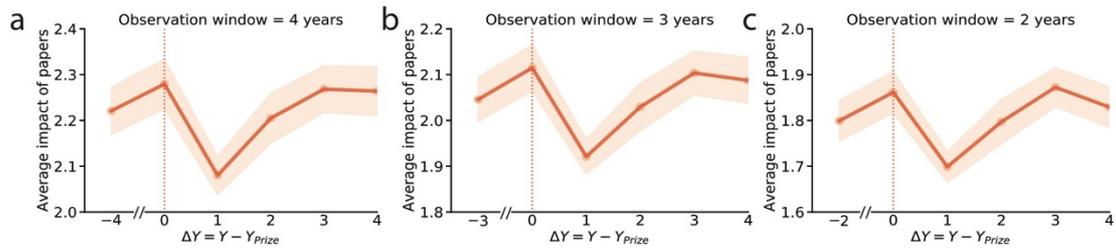

**Fig. S20 | Average impact per paper as a function of career years. a** We set the observation window as 4 years and calculate the average impact of papers based on the 4-year citation counts. For each laureate, we exclude all works where the impact measure includes post-Prize citations and compare the average impact of papers published by the laureates in the 4th year before and each of the four years after winning the Nobel. The solid line indicates the average across all laureates in our sample four years before and after the Nobel, as well as the prize winning year, with the shaded area denoting the standard error of the mean. **b-c** Average impact per paper as a function of career years with the observation windows as 3 years and 2 years respectively.

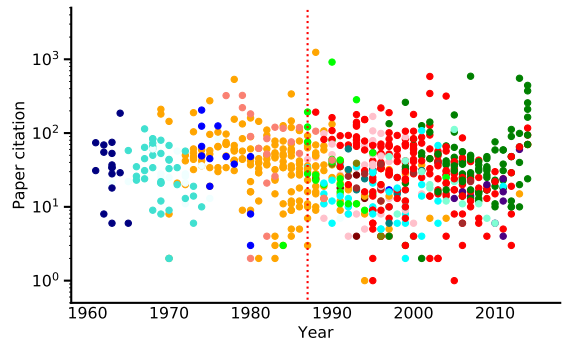

**Fig. S21 | The publication history of Nobel laureate Jean-Marie Lehn by color coding each paper according to the related topic community based on the Infomap community detecting algorithm.** The X-axis shows the time series of all the papers published by Nobel laureate Jean-Marie Lehn, and the Y-axis shows the citation for each paper. Each paper is represented by a point and the color corresponds to the topic community in the co-citing network.

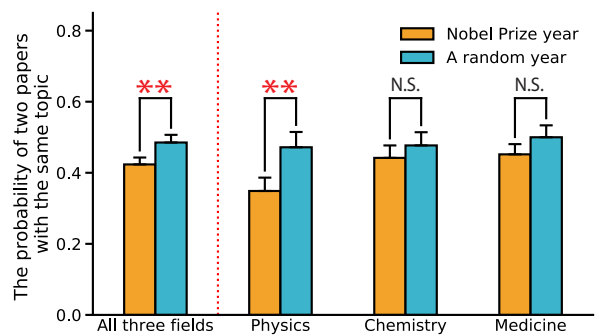

**Fig. S22 | Comparison of the probability of two papers belonging to the same topic within 4 years before and after the reception of the Prize and a random year.** The probability is significant lower after winning the Prize, suggesting Nobel laureates tend to shift research topic

after winning the Nobel. The change is significant for physics, while there is no significant difference for chemistry and medicine. *** p<0.01, ** p<0.05, * p<0.1 and N.S. (not significant) for p>0.1. Error bars represent the standard errors of the mean.

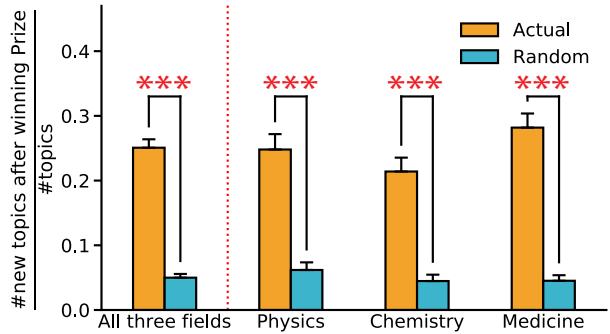

**Fig. S23 | Comparison of the likelihood of laureates shifting to a new topic after winning Prize with random.** We measured the likelihood of laureates shifting to a new topic after winning Prize, $\frac{\#new\ topics\ after\ winning\ Prize}{\#topics}$. Then, we shuffled the topic of the works and repeated the measurement as a null model, finding that the laureates are much more likely to shift to a new topic after winning Prize than random. The observation is significant for all three fields. *** p<0.01, ** p<0.05, * p<0.1 and N.S. (not significant) for p>0.1. Error bars represent the standard errors of the mean.

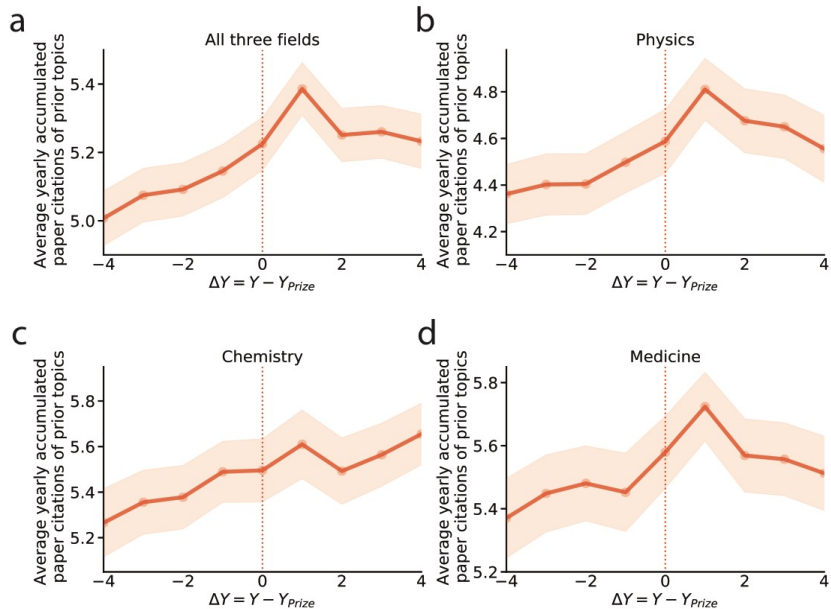

**Fig. S24 | Average yearly accumulated citation counts of prior topic papers as a function of career years.** We focus on laureates' papers that belong to the prior Prize topics, and measure its yearly accumulated citation counts before and after the Prize. The solid line indicates the average across all laureates in our sample four years before and after the Nobel, as well as the prize winning year, with the shaded area denoting the standard error of the mean.

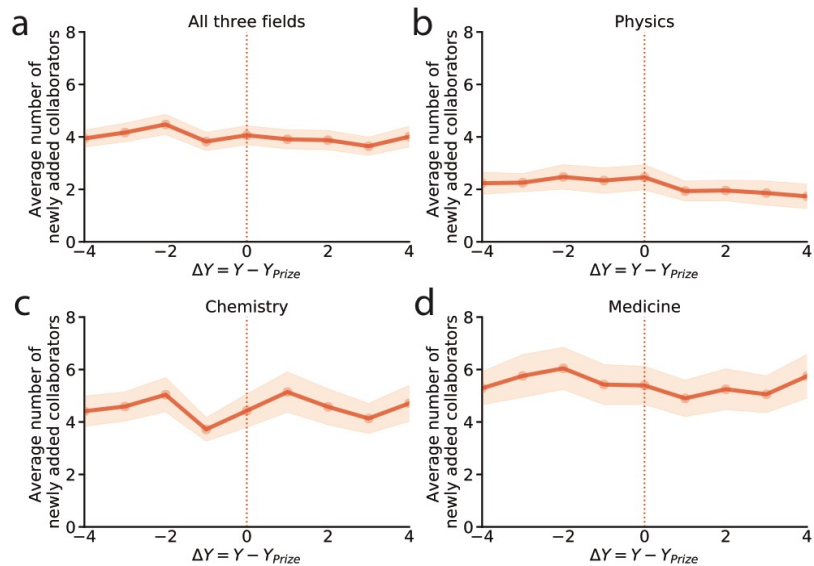

**Fig. S25 | Average number of new collaborators before and after the Prize.** For each laureate, we calculate the new collaborators each year. The solid line indicates the average across all laureates in our sample four years before and after the Nobel, as well as the prize-winning year, with the shaded area denoting the standard error of the mean.

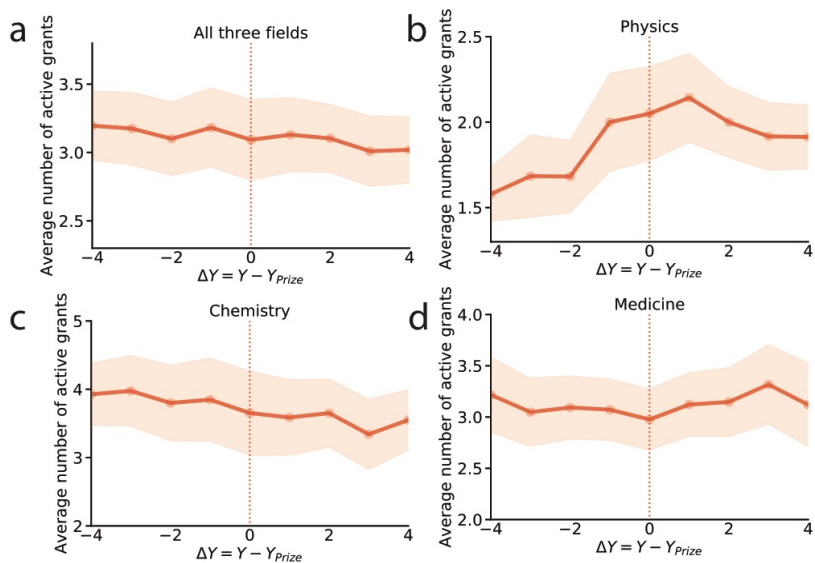

**Fig. S26 | Funding opportunities changing before and after the Prize. a-d** Average number of active grants as a function of career years. We only focus on those laureates who participated in at least one active grant project both within 5 years before and after the Prize, thus resulting in 25 Physics laureates, 46 Chemistry laureates and 43 Medicine laureates. The year when the Prize is given is marked as 0. For each laureate, we calculate the active grant number he/she involved in each of the four years before and after the Nobel, as well as the prize-winning year. The solid line indicates the average across all laureates in our sample, with the shaded area denoting the standard error of the mean.